\documentclass[sigconf]{acmart}
\AtBeginDocument{%
  \providecommand\BibTeX{{%
    \normalfont B\kern-0.5em{\scshape i\kern-0.25em b}\kern-0.8em\TeX}}}

\copyrightyear{2024}
\acmYear{2024}
\setcopyright{rightsretained}
\acmConference[ASSETS '24]{The 26th International ACM SIGACCESS Conference
on Computers and Accessibility}{October 27--30, 2024}{St. John's, NL, Canada}
\acmBooktitle{The 26th International ACM SIGACCESS Conference on Computers
and Accessibility (ASSETS '24), October 27--30, 2024, St. John's, NL, Canada}
\acmDOI{10.1145/3663548.3675611}
\acmISBN{979-8-4007-0677-6/24/10}

\usepackage{algorithm}
\usepackage{algpseudocode}
\begin{document}

\title[ChartA11y]{ChartA11y: Designing Accessible Touch Experiences of Visualizations with Blind Smartphone Users}

\author{Zhuohao (Jerry) Zhang}
\affiliation{%
 \institution{University of Washington}
 \city{Seattle}
 \state{Washington}
 \country{USA}
 }
 
\author{John R. Thompson}
\affiliation{%
 \institution{Autodesk Research}
 \city{Atlanta}
 \state{Georgia}
 \country{USA}
 }
 
\author{Aditi Shah}
\affiliation{%
 \institution{Microsoft Corporation}
 \city{Redmond}
 \state{Washington}
 \country{USA}
 }
 
\author{Manish Agrawal}
\affiliation{%
 \institution{Microsoft Corporation}
 \city{Redmond}
 \state{Washington}
 \country{USA}
 }
 
\author{Alper Sarikaya}
\affiliation{%
 \institution{Microsoft Corporation}
 \city{Redmond}
 \state{Washington}
 \country{USA}
 }
 
\author{Jacob O. Wobbrock}
\affiliation{%
 \institution{University of Washington}
 \city{Seattle}
 \state{Washington}
 \country{USA}
 }
 
\author{Edward Cutrell}
\affiliation{%
 \institution{Microsoft Research}
 \city{Redmond}
 \state{Washington}
 \country{USA}
 }
 
\author{Bongshin Lee}
\authornote{corresponding author}
\authornotemark[0]
\affiliation{%
 \institution{Yonsei University}
 \city{Seoul}
 \country{Republic of Korea}
 }

\renewcommand{\shortauthors}{Zhang et al.}

\begin{abstract}

We introduce ChartA11y, an app developed to enable accessible 2-D visualizations on smartphones for blind users through a participatory and iterative design process involving 13 sessions with two blind partners. We also present a design journey for making accessible touch experiences that go beyond simple auditory feedback, incorporating multimodal interactions and multisensory data representations. Together, ChartA11y aimed at providing direct chart accessing and comprehensive chart understanding by applying a two-mode setting: a semantic navigation framework mode and a direct touch mapping mode. By re-designing traditional touch-to-audio interactions, ChartA11y also extends to accessible scatter plots, addressing the under-explored challenges posed by their non-linear data distribution. Our main contributions encompass the detailed participatory design process and the resulting system, ChartA11y, offering a novel approach for blind users to access visualizations on their smartphones.
\end{abstract}

\ccsdesc[500]{Human-centered computing~Human computer interaction (HCI)}
\ccsdesc[500]{Human-centered computing~Accessibility technologies}
\ccsdesc[500]{Human-centered computing~Visualization systems and tools}

\keywords{Assistive Technology, Smartphone, Touchscreen Experience, Data Visualization, Sonification, Multimodal Interaction, Blind Users, Participitory Design}

\maketitle

\section{Introduction}

Data visualizations play a crucial role in understanding and exploring complex datasets, and in gaining insights such as trends, correlations, and outliers. Despite their effectiveness for sighted users, the inherently visual nature of data visualizations poses significant accessibility barriers for blind and low-vision individuals (BLVIs). Screen readers, which serve as essential assistive technologies for BLVIs to access digital content, are not optimally designed to interpret and convey the multidimensional aspects of charts. This incompatibility is amplified by the complexity of data visualizations, which often render as multiple data trends, clusters, and correlations, making navigation and comprehension through auditory feedback alone particularly challenging. Moreover, more complex chart types, such as scatter plots, further complicate accessibility due to their non-linear and spatially distributed data points. 

The field of accessible visualization research has witnessed substantial efforts from the community. Traditional accessibility guidelines \cite{wai-aria} often advise visualization authors to provide alternative textual or tabular representations of graphical data. This method theoretically enables screen reader users to access and analyze data, potentially even applying statistical tools. Yet, this approach demands significant technical expertise and imposes a heavy cognitive load on BLVIs. 
The goal of ensuring that BLVIs can access data as quickly and intuitively as sighted individuals do when they glance at a visualization highlights the need for simpler and more effective methods of accessing.

In response, researchers have created numerous approaches including different navigational structures that allow screen reader users to interact with charts through detailed aspects such as insights, axes, data points, and filters, often in customizable ways \cite{thompson_chart_2023, jones_customization_2024, zong_rich_2022}. Among the investigated modalities, touch-based experiences \cite{butler_technology_2021} have attracted interest for their capacity to improve accessibility. However, the focus has predominantly been on simplistic touch-to-audio feedback mechanisms, raising a critical question: Can screen reader users effectively process complex information in a two-dimensional (2-D) space without the benefit of hand-eye coordination? Our investigation takes this discourse further by examining the interaction of BLVIs and charts with touchable elements, and identifying previously unrecognized challenges. For example, we noted that even with immediate feedback like audio and haptics, BLVIs still frequently face challenges when following visual traces like the curve of a line, the boundary of a data cluster, or simply horizontal directions within a chart. The set of under-explored challenges makes it difficult for BLVIs to discern underlying trends, shapes, or relationships among data points distributed across a 2-D space.

To address the challenges inherent in creating accessible touch experiences for visualizations, we conducted an iterative participatory design approach, collaborating with two blind smartphone users across 13 sessions over eight months. This method diverged from conventional user studies, where blind participants are typically only introduced to formative or summative prototypes~\cite{ladner_design_2015}. Instead, our collaborators were integrally involved in every step of the design and development process from the outset. We dived into the specific challenges faced by BLVIs when using standard touch-to-audio methods for accessing data visualizations. 

Through iterative design and testing, we developed \textit{ChartA11y}, a system offering interaction techniques for making line charts, bar charts, and scatter plots accessible. Ultimately, our blind collaborators could independently navigate complex charts with ChartA11y, extracting data insights from multi-series scatter plots comparably to how sighted users do so. Our contributions in this work include: 

\begin{enumerate}
\item ChartA11y, a functional system with multiple interaction techniques that facilitate non-visual access to complex charts on smartphones---unachieved by prior research, which mostly focused on simpler visualizations like single-series scatter plots with clear patterns. 

\item A design journey, which highlights how various assistive techniques can initially fall short in terms of their effectiveness for BLVIs despite our initial expectation that they would benefit our target users.
\end{enumerate}

\section{Related Work}

Our work is informed by existing research on (1) accessible visualization experiences for BLVIs and (2) approaches of multisensory data representation and multimodal interactions in assistive technology design.

\subsection{Accessible Visualization Experiences}

Recent efforts within the HCI, accessibility, visualization, and AI research communities have been directed towards rendering traditionally inaccessible visual content accessible. A number of studies \cite{kim_accessible_2021, sharif_understanding_2021, joyner_visualization_2022, choi_visualizing_2019, lazar_what_2007} have investigated the current landscape of chart accessibility, identifying significant knowledge gaps and socio-technical challenges within this domain. Fan et al. \cite{fan_accessibility_2023} pinpointed prevalent accessibility issues in online visualizations, which are often crucial for conveying information to BLVIs. Lundgard et al. \cite{lundgard_sociotechnical_2019} emphasized that contemporary technologies may inadvertently introduce barriers, advocating for clear communication of accessibility needs and the inclusion of individuals with disabilities as equal stakeholders in the design process. Marriott et al. \cite{marriott_inclusive_2021} noted that despite advances, understanding of how to effectively support accessible visualizations lags. Elavsky et al. \cite{elavsky_how_2022} proposed a set of heuristics for visualization authors aimed at transcending minimal accessibility standards to create genuinely usable experiences.

In the realm of designing digital and practical assistive technologies, significant strides have been made in different venues. Support from mainstream screen readers such as NVDA \cite{noauthor_nv_nodate}, JAWS \cite{noauthor_jaws_nodate}, iOS VoiceOver, and Android TalkBack enables BLVIs to navigate web pages and mobile applications containing visual content. Despite these advancements, the majority of online-based visualizations remain largely inaccessible to screen readers due to the different rendering methods. In the meantime, commercial efforts such as Audio Graphs \cite{noauthor_audio_nodate} and Highcharts \cite{noauthor_interactive_nodate} have endeavored to offer accessible alternatives. 

Within the academic community, innovative projects have aimed to create more engaging and interactive visualization experiences. Early initiatives like the iGRAPH-Lite system \cite{ferres_evaluating_2013} facilitated chart navigation and caption generation via keyboard inputs. Zong et al. \cite{zong_rich_2022} structured screen reader interactions with chart elements in a hierarchical manner. Sharif et al.'s VoxLens \cite{sharif_voxlens_2022} and its subsequent extension \cite{sharif_understanding_2023} enhanced online visualization accessibility through voice-activated commands for ``Q\&A'' and ``drill-down'' information retrieval interactions.
Additionally, research has been conducted on optimizing the textual output received by BLVIs. Jung et al. \cite{jung_communicating_2022} explored the necessary levels of textual support. Sharif et al. \cite{sharif_conveying_2023} looked into methods for communicating uncertainty in visualizations, discovering a preference among users for statistical information to be presented in plain language and in a detailed manner. Leveraging the Olli accessible visualization toolkit \cite{blanco_olli_nodate}, Jones et al. \cite{jones_customization_2024} enabled speech output customizations to simplify the identification and recall of chart information.

Within this landscape, AI has increasingly been employed to automatically extract content and insights from visualizations. Studies have introduced interactive and semi-automatic systems for data extraction from chart images \cite{savva_revision_2011, jung_chartsense_2017}. Lundgard and Satyanarayan \cite{lundgard_accessible_2022} conceptualized how natural language descriptions could semantically articulate the content of visualizations. Battle et al. developed Beagle \cite{battle_beagle_2018} for automatic insight extraction from online visualizations. Several works have applied AI solutions, including transformer models \cite{obeid_chart--text_2020} and deep neural networks \cite{kim_information_2021, kim_multimodal_2018, alam_seechart_2023}, to summarize and generate natural language descriptions of charts. 

\subsection{Multisensory Data Representation and Multimodal Interactions}
The integration of multisensory data representations stands as a significantly notable practice within accessibility research, encompassing a broad spectrum of assistive technology designs beyond just data visualizations. Utilizing auditory cues and sonifications, researchers have significantly advanced the accessibility of diverse digital content for visually impaired users \cite{lee_collabally_2022, microsoft-soundscape, white_toward_2008, bandukda_audio_2020, mendes_collaborative_2020, goncalves_playing_2020, hoque2023accessible}. Specifically in the context of visualizations, early studies have assessed the efficacy of auditory scatter plots in conveying the direction and magnitude of correlations \cite{flowers_cross-modal_1997}. Further explorations have focused on evaluating the usability and effectiveness of sensory substitutions \cite{chundury_towards_2022} and sonified feedback mechanisms \cite{sharif_what_2022}. Prior work such as iSonic by Zhao et al. \cite{zhao_data_2008} offers support for visually impaired users in navigating georeferenced data through enriched auditory and speech output. Similarly, Seo et al. introduced the MAIDR system \cite{seo_maidr_2024} that extends modalities to include braille, alongside conventional sonification and textual descriptions. Infosonics by Holloway et al. \cite{holloway_infosonics_2022} integrates diverse audio tracks to facilitate the interpretation of infographic trends. More recently, Zong et al. \cite{zong_umwelt_2024} present visualization, sonification, and textual descriptions as equal mediums to enhance accessibility of charts.

Another promising venue to substitute traditional textual feedback in accessible visualization and other types of digital media \cite{scoy_auditory_2005, gorlewicz_design_2020, brown_design_2003, shi_designing_2019, mcgookin_soundbar_2006, mullenbach_surface_2013, blenkhorn_using_1998} is to use multimodal interactions, like touch, haptics, and vibrotactile feedback \cite{butler_technology_2021, chundury2023tactualplot}. Chundury et al. proposed TactualPlot \cite{chundury2023tactualplot}, which spatializes data as sound using sensory substitution. Noteworthy contributions include the work by Palani et al. \cite{antona_haptic_2018}, which proposed guidelines for depicting oriented lines on touchscreen devices, leveraging vibrotactile stimuli to convey information. Awada et al. \cite{awada_towards_2012} explored protocols that enable BLVIs to identify geometric shapes on vibrating touchscreen interfaces. Furthermore, recent work by Moore et al. \cite{moore_spatial_2024} investigated spatial-audio enhanced visualizations on touch screen devices to support BLVIs in learning data trends efficiently. 

Given the rich set of works from various research communities, ChartA11y sets out to explore a distinct aspect. Given the widespread adoption of smartphones, blind individuals are increasingly likely to interact with charts on their mobile devices. This shift presents unique opportunities for improving the accessibility of visualizations, leveraging the inherent multimodal capabilities of these devices. However, this transition is accompanied by a new set of challenges. Our aim is to systematically identify these challenges and, through an iterative design process, develop an effective solution to enable accessible touch experience on smartphones.

\section{Participatory Design Method}

In this section, we introduce the participatory design method employed in our research, setting the stage for a detailed presentation of the ChartA11y system design in the subsequent section. This overview serves to acquaint readers with the foundational approach underpinning our research, prior to diving into the intricacies of the design journey itself.

Our research involved two blind co-authors who contributed to the iterative design process over an eight-month period and 13 co-design sessions. Unlike traditional research paradigms, where blind participants may engage solely as subjects in the final stages, our collaborators were integral to our iterative design and development from the outset, responding to calls from the community \cite{ladner_design_2015, lundgard_sociotechnical_2019}. The accessibility community has also been widely adopting this co-design approach with BLVIs \cite{thompson_chart_2023, morrison_social_2021, zhang_developing_2023, van_rooyen_shape2vibe_2024, feng_designing_2016}. This inclusive approach ensures that system creation is closely aligned with the actual needs and experiences of BLVIs, moving beyond passive participation to active co-creation.

Each session, lasting between 60 to 90 minutes, commenced with an update on the system's progress, incorporating insights and feedback from the previous meeting. Subsequently, the sessions focused on hands-on interaction with the system, where blind collaborators were introduced to new features based on their growing familiarity with the system. To differentiate learning from testing, we employed different charts for tutorials and tasks (eight charts in total), ensuring that the collaborators' ability to perform interpretive tasks was assessed without prior context. For example, both partners were asked to identify potential correlations between $X$ and $Y$ values, distinguish clusters that are obvious to sighted users, spot outliers in a scatter plot, and compare data distributions from different series.

\section{ChartA11y Prototype}
\label{sec:charta11y}

We present the final design iteration of ChartA11y, including its implementation and interaction techniques. We then describe the design journey and insights that led to ChartA11y in Section \ref{sec:journey}.

\subsection{ChartA11y Overview}
\begin{figure*}
    \centering
    \includegraphics[width=0.9\textwidth]{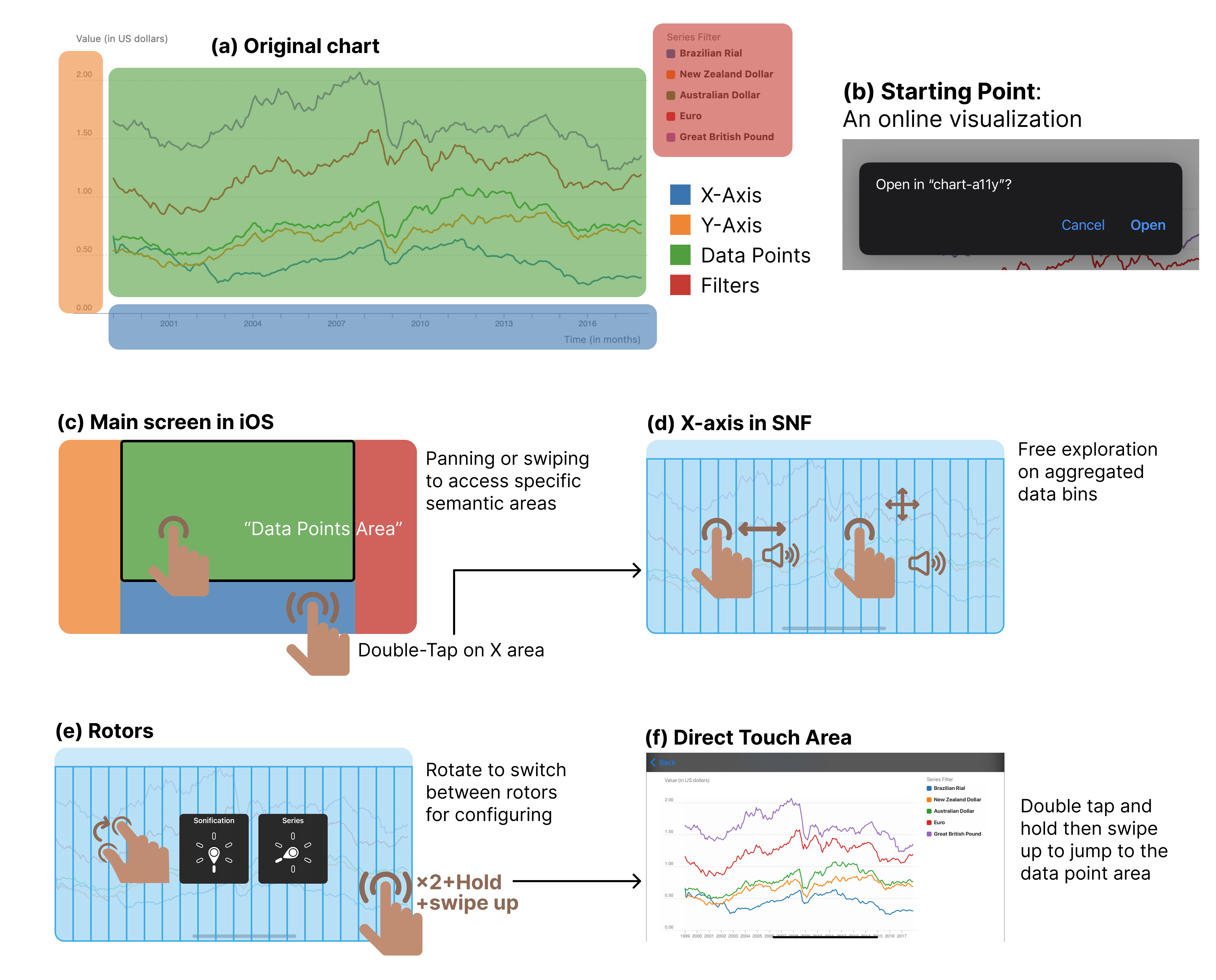}
    \caption{The design of the Semantic Navigation Framework mode, including (a) the original chart rendered on a web page, (b) the starting point of how blind users open charts in ChartA11y, and (c-f) the example scenes of using SNF mode.}
    \Description{A complex figure consisting of multiple sections to convey the design of semantic navigation framework, including (a) the original chart of a multi-series line chart rendered on a web page with four different colors to represent the different chart areas to be semantically mapped onto the touch device, (b) the starting point of how blind users open charts in ChartA11y, where users click a button to jump to and open this chart in ChartA11y, and (c-f) the example scenes of using SNF, with 4 steps from the beginning of opening ChartA11y to the users wanting to explore details and going to the direct touch mapping mode.}
    \label{fig:snf}
\end{figure*}

ChartA11y is built as an Apple iOS\footnote{Our use of iOS devices stems from the prevalence of iPhones and their VoiceOver touch-based screen-reading software among BLVIs.} mobile application that opens an existing chart on the Web in an accessible format (\autoref{fig:snf}b). The rendering of charts is powered by the open-sourced Chart Reader software \cite{thompson_chart_2023}, a web-based accessibility engine that takes in a dataset in CSV format and renders an SVG chart to a web page. By leveraging this technology, ChartA11y effectively simulates the experience of smartphone users encountering visualizations online, enabling these charts to be opened within ChartA11y. For simplicity, we assume that the underlying source data for visualizations are available. Approaches to overcoming this assumption, which are well explored in machine learning and computer vision~\cite{sreevalsan-nair_tensor_2020, ma_towards_2021, liu_data_2019}, are left to future work. (Similar assumptions have also been made by recent work in touch accessibility~\cite{li_toucha11y_2023, liang_brushlens_2023}.)

\subsection{ChartA11y Interactions}
\label{sec:interaction}
ChartA11y is distinguished by its integration of two innovative modes: (1) the Semantic Navigation Framework (SNF) mode and (2) the Direct Touch Mapping (DTM) mode. The two modes work together to provide a comprehensive understanding of charts. The SNF mode is engineered to offer a VoiceOver-like experience, semantically organizing the chart's navigational structure within the mobile application. This mode allows users to intuitively explore the visualization \textit{as if} the chart were physically mapped onto the touch display. Conversely, the DTM mode grants blind users direct access to data points, simulating a tactile representation of the chart on the screen. ChartA11y enables users to easily switch between these modes, maintaining awareness of their position within the chart or in the semantic navigational structure. In the following sections, we describe the detailed interactions and corresponding multisensory data representations of ChartA11y.

\subsubsection{Semantic Navigation Framework (SNF)}

ChartA11y provides a semantic navigation framework (\autoref{fig:snf}) that renders different components of the chart semantically onto the touch display for an accessible touch experience.

\paragraph{Fundamental Interaction Logic}
The SNF's central interaction is ``finger-reading''~\cite{kane_slide_2008,zhang_a11yboard_2023} without encountering any blank areas, as the entire touch space is semantically populated with chart elements. This design ensures that any touch by users always elicits a meaningful response, facilitating a continuous and intuitive exploration of the chart's structure and data points \cite{blanch_semantic_2004}. Chart elements are rendered as accessible, rectangular views, ensuring that navigation through different parts of the chart is straightforward and effective, as shown in \autoref{fig:snf}(c)-(f). As users shift their focus to a new element by panning or swiping left or right, immediate feedback is provided via VoiceOver announcements or sonification audio tones, depending on the selected audio mode. 

The navigation begins with an overview of the chart (\autoref{fig:snf}c), presented as a full-screen view, from which users can dive into four distinct semantic zones: the $X$ axis, $Y$ axis, filters, and data points. Each zone further unfolds into lists of data bins or cells, tailored to the specific type of chart being explored. For clarity in subsequent discussions, we define ``data bins'' and ``data cells'' as such: Data bins are used to describe aggregated data segments on the $X$ or $Y$ axis, where data is grouped based on ranges. Data cells refer to discrete sections within scatter plots, delineated by intersecting $X$ and $Y$ values, facilitating a granular analysis of data point distributions. 

\begin{figure*}
    \centering
    \includegraphics[width=.8\textwidth]{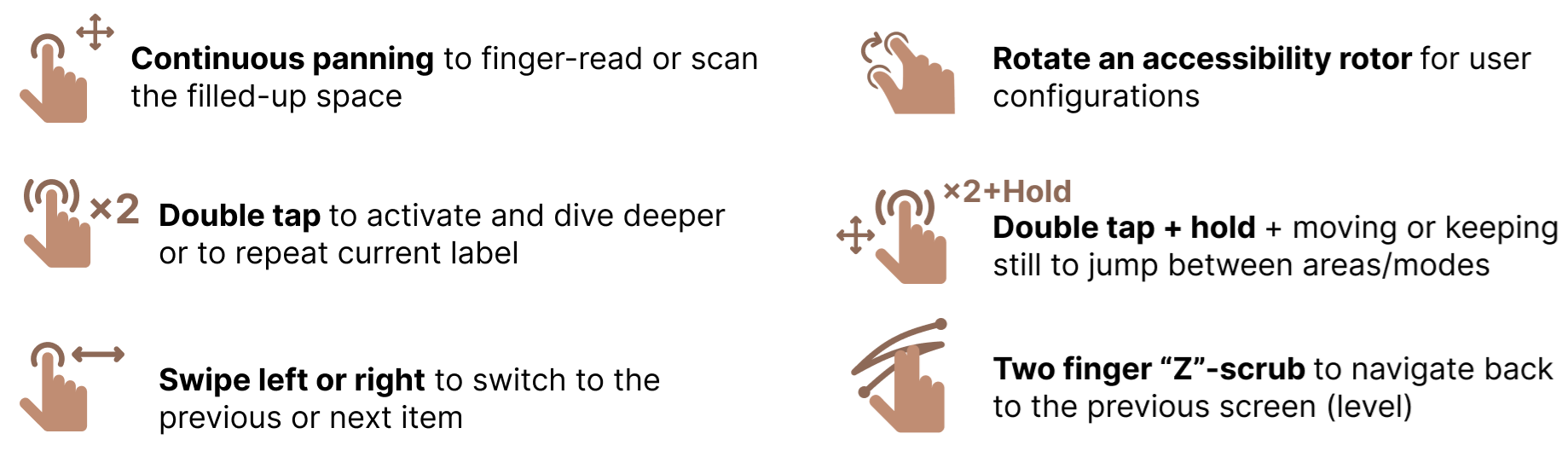}
    \caption{The design of touch and gestures in the Semantic Navigation Framework, with most similar to VoiceOver's gestures to give users a similar navigation experience.}
    \Description{A series of touch and gesture icons coupled with textual explanations. (1) Continuous panning to finger-read or scan the filled-up space, (2) Double tap to activate and dive deeper or to repeat current label, (3) Swipe left or right to switch to the previous or next item, (4) Rotate an accessibility rotor for user configurations, (5) Double tap + hold + moving or keeping still to jump between areas/modes, (6) Two finger “Z”-scrub to navigate back to the previous screen (level).}
    \label{fig:gesture}
\end{figure*}

\paragraph{VoiceOver Touch and Gesture}
The SNF leverages the intuitive touch and gesture controls familiar to blind smartphone users---specifically, VoiceOver gestures---to facilitate access to chart elements. These gestures (\autoref{fig:gesture}) include swiping to navigate through items, double-tapping to drill down to deeper layers of data, executing a two-finger ``$z$''-scrub to return to a previous navigation level, and employing touch to directly select a particular data point. To accommodate instances where a navigation level contains a high density of items, a paging mechanism is implemented, which ensures that each data item is allocated a minimum width or height sufficient for finger recognition. Consequently, users can employ a three-finger swipe to move seamlessly between pages. 

\paragraph{Series Switching Rotor}
In iOS VoiceOver, rotor is an accessibility widget that enables convenient mode switching. ChartA11y introduces custom accessibility rotors for enhanced navigation, including a ``series'' rotor enabling users to cycle through data series or access an overview summary. This feature, coupled with the ability to toggle specific series in the filter area, enriches the data exploration experience. To use the rotor, BLVIs can rotate two fingers on the screen as if they are turning a dial. After they choose an option, BLVIs can flick their finger up or down on the screen to select different options. In our case, they can switch between all data series.

\paragraph{Sonification Toggle Rotor}
A dedicated rotor toggles sonification on or off, replacing visual labels with audio tones. These tones vary intelligently across different scenarios and data series, offering users a nuanced auditory representation of the data. The sonification settings are adaptable, catering to different investigation needs and personal preferences.

\begin{figure*}
    \centering
    \includegraphics[width=.9\textwidth]{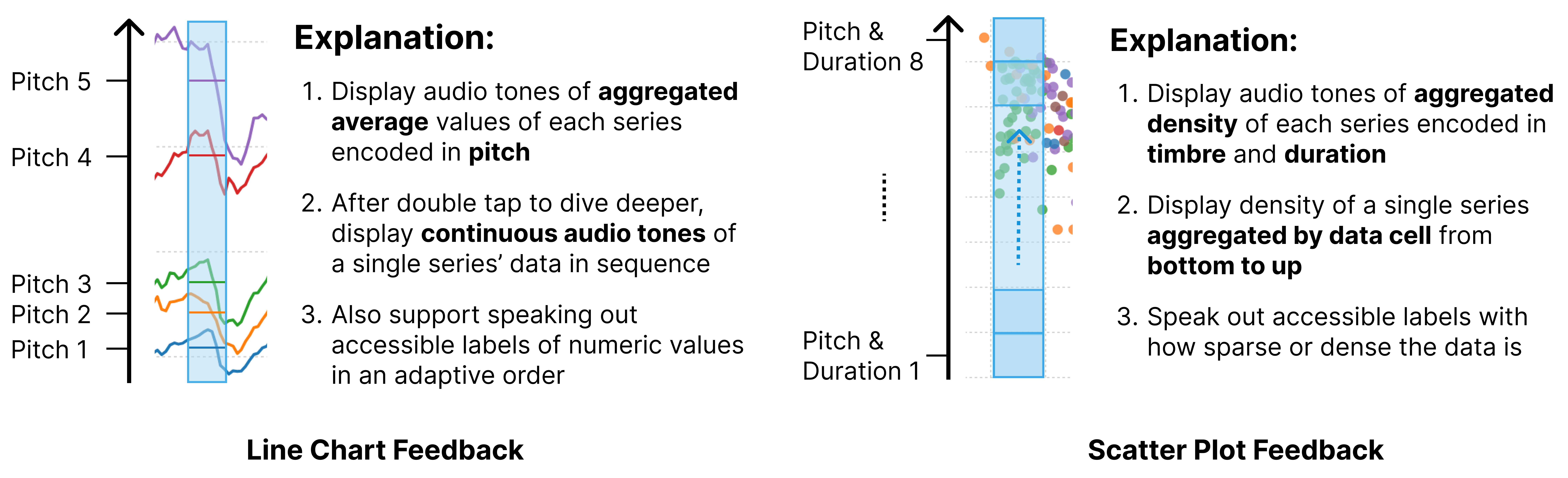}
    \caption{The design of auditory feedback in Semantic Navigation Framework mode, including two examples of feedback in line charts and scatter plots.}
    \Description{The perceived sonification and speech feedback description, with two examples of line chart and scatter plot. Explanation for line chart: Display audio tones of aggregated average values of each series encoded in pitch; After double tap to dive deeper, display continuous audio tones of a single series’ data in sequence; Also support speaking out accessible labels of numeric values in an adaptive order. Explanation for scatter plot: Display audio tones of aggregated density of each series encoded in timbre and duration; Display density of a single series aggregated by data cell from bottom to up; Speak out accessible labels with how sparse or dense the data is.}
    \label{fig:feedback}
\end{figure*}

\paragraph{Sonification and Encodings}
Sonification in ChartA11y (\autoref{fig:feedback}) is carefully designed to reflect various data attributes. For line and bar charts, the sonification strategy employs pitch variation to represent changes in data values, which was widely adopted in prior research and products \cite{thompson_chart_2023, chundury_towards_2022, noauthor_audio_nodate}, allowing users to perceive trends and fluctuations in the data as they move their fingers across the chart. This method ensures that users can follow the progression of data points through a continuous auditory experience, where higher pitches correspond to higher values and vice versa, enabling an intuitive understanding of the data's highs and lows. 

In scatter plots, however, the encoding strategy shifts from focusing on pitch only to also incorporating tone duration to convey the density of data points within a chart's grid cell. This adjustment reflects the unique challenge of interpreting scatter plots, where the spatial distribution of points conveys significant information. With this enhanced encoding, users can either explore individual data cells or aggregated data bins in different navigational levels. When engaging with data cells, the sonification experience is akin to freely exploring the 2-D space with one's finger, where each contact with a data cell triggers a distinct audio tone. On the other hand, navigating through data bins involves a sequential sonification approach, where users encounter a series of audio tones that represent each cell within a bin, moving methodically from one to the next. For instance, in a scenario where a user examines a data bin along the $X$-axis containing nine data cells, they would be presented with nine sequential audio tones. These tones vary from ``numb'' sounds, signifying the absence of data points in a cell, to meaningful musical notes that differ in pitch and duration, indicating the presence and density of data points.

\paragraph{Accessible Narration Design}

The design of accessible labels for each element is a critical component in ChartA11y's navigation framework. These labels follow a conventional structure by reporting the $X$ and $Y$ values, along with any related filter information. Additionally, we have enriched this basic model with innovative features to create a more intuitive and informative interaction.

First, we introduced a location-aware and adaptive data narration feature. It dynamically adjusts the narration based on the user's interaction with the charts. When a user touches a new position on the screen or swipes to navigate through data items, ChartA11y discerns whether the user has landed on a new position or is moving to an adjacent item. For a new position, the accessible label prioritizes positional information by announcing the $X$ value first, catering to the user's need to understand their new location. Conversely, when navigating to a neighboring data item, the system shifts focus to announce the $Y$ value first, recognizing that the user, familiar with their current position, seeks detailed data information next. This adaptive approach ensures that users receive the most relevant information based on their navigation pattern.

\begin{figure*}
    \centering
    \includegraphics[width=\textwidth]{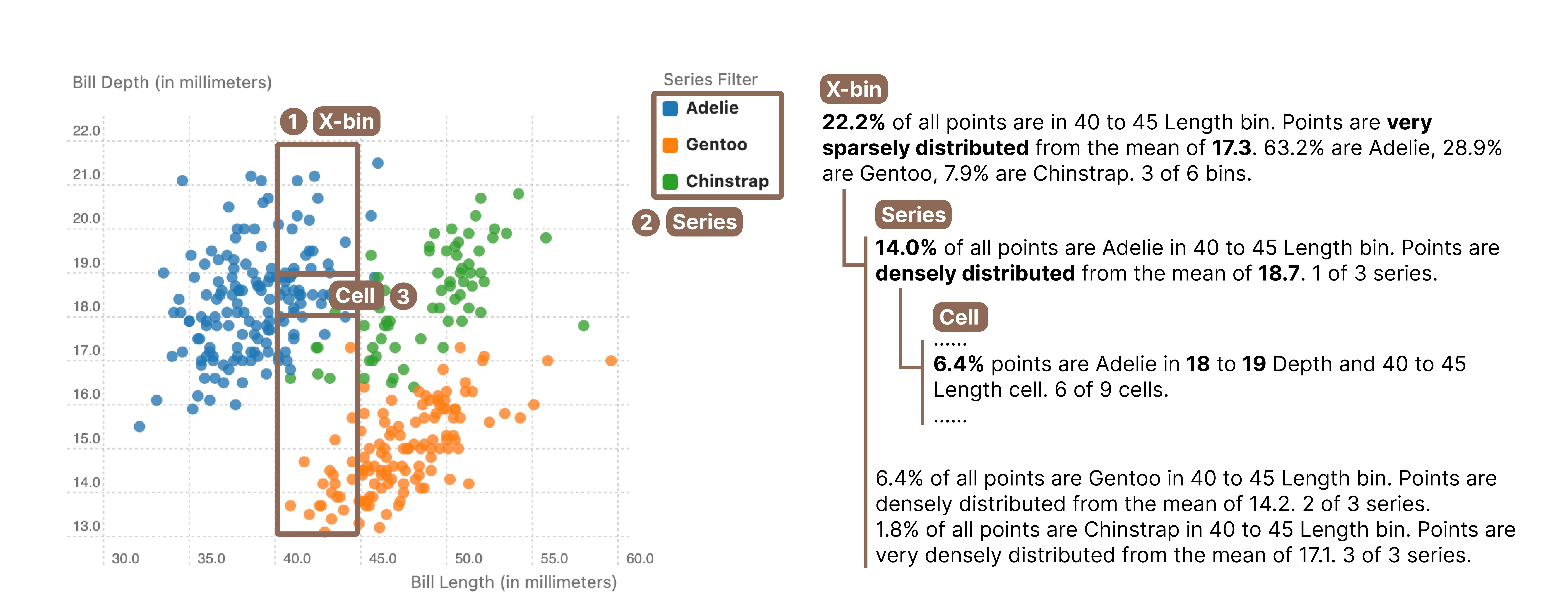}
    \caption{An example of accessible narration design in scatter plots, including the bin, series, and cell levels.}
    \Description{The example of accessible narration design in scatter plots, including the bin, series, and cell levels. The left side shows the original chart of a penguin dataset, and the right side shows the three different levels of narration.}
    \label{fig:label}
\end{figure*}

Second, our iterative design process has refined how we report data bins in line and bar charts, or cells in scatter plots. An illustrative example of this can be seen in the accessible narration for a scatter plot (\autoref{fig:label}), where the label first indicates the percentage of data points within the current bin, followed by a qualitative description of the data distribution within that range. This description varies from very sparsely distributed to very densely distributed, providing users with a clear explanation of the data's spatial arrangement. Going deeper, users can explore specific data series within a bin and access detailed distribution information about data cells. This layered approach to data narration ensures that users can both grasp the overall data landscape and explore the minutiae of data distributions and relationships. Together with the sonification toggle introduced above, this integration allows for smooth switching between descriptive feedback and auditory feedback for detailed inforation or data trends and  distributions, respectively.

\subsubsection{Direct Touch Mapping (DTM)}

\begin{figure*}
    \centering
    \includegraphics[width=\textwidth]{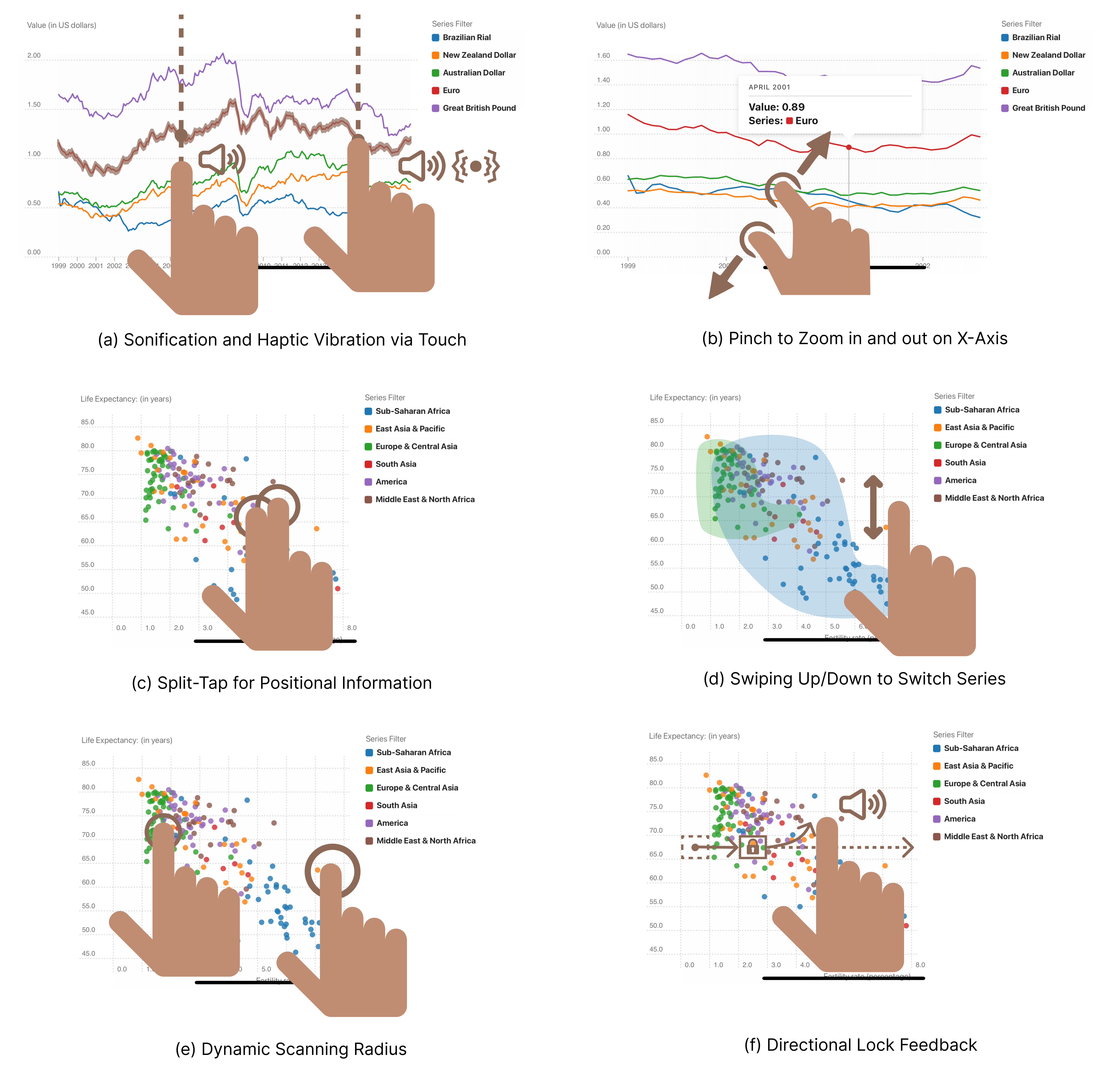}
    \caption{The design of Direct Touch Mapping mode, including different features to enable BLVIs to directly and effortlessly access visual elements in chart.}
    \Description{The design of Direct Touch Mapping mode, including six different sub-figures showing the exact interactions.}
    \label{fig:dtm}
\end{figure*}

Beyond the Semantic Navigation Framework (SNF), our design process also iterated and adapted a distinct interaction mode called Direct Touch Mapping (DTM). This mode diverges from semantic navigation by offering blind users direct access to the original chart elements. The direct touch mapping was achieved by enabling ``direct touch'' on an accessibility view in Apple iOS, allowing us to provide immediate feedback. The interactions in DTM are tailored to manage the cognitive load of blind users, ensuring the dense 2-D information remains accessible and affordable without overwhelming them. In the following subsections, we describe the specific interactions in DTM, and how we integrated it with the SNF, allowing blind users to fluidly switch between the two without losing their sense of position.

\paragraph{Fundamental Interaction Logic}
The core interaction principle in DTM continues to leverage speech or audio tone feedback in response to a user's touches. However, our approach advances beyond the conventional touch-to-audio methodology by incorporating additional modalities and considerations in our design. The detailed design of DTM differs between line or bar charts and scatter plots, as described below.

\paragraph{Line and Bar Charts}
For line and bar charts, which typically display data linearly along the X-axis, ChartA11y provides audio feedback correlating with the user's direct touch position projected on the $X$-axis. This interaction mimics a slider mechanism, supplemented by vibration feedback upon contact with actual visual elements, like lines or bars (\autoref{fig:dtm}a). This tactile feedback offers users a more engaged understanding of data distribution, moving beyond passive auditory sonification to actively tracking data trends. Audio tone adjustments are calibrated to the speed of the user's finger movement~\cite{kane_slide_2008}, employing a throttle technique to ensure the consistency of meaningful audio feedback, regardless of how swiftly the finger moves.

Zoom functionality, achieved through pinching gestures (\autoref{fig:dtm}b), allows users to focus on specific data segments. For example, in a densely packed three-year COVID-19 data chart, users can zoom into a particular year or month for a closer look. This feature, akin to that in the SNF, also permits users to switch between data series by swiping up or down.

ChartA11y enriches data exploration by enabling users to access detailed information beyond simple data point interactions. The design is facilitated through a split-tap gesture \cite{kane_slide_2008}, where users tap with a second finger while the first remains on the screen, providing immediate access to positional details (\autoref{fig:dtm}c). Additionally, direct interaction with the $X$ and $Y$ axes offers users precise information on axis values and labels, further enhancing users' navigational and interpretive capabilities within the chart. These features collectively support a more comprehensive and intuitive exploration of data visualizations for blind users.

\paragraph{Scatter plots}
Scatter plots are complex and require a tailored set of interactions mindful of blind users' cognitive load and working memory.
Given that scatter plots often feature densely populated data points, ChartA11y employs a radius-based scanning technique. The system treats a fingertip as a scanning window; as users move their finger, the ``window'' encounters data points, triggering haptic feedback via the device's vibration engine for each data point contacted. This mechanism emulates tactile interfaces, providing a continuous haptic experience mirroring the sensation of touching raised data points. The scanning window's size dynamically adjusts based on the local density of data points, ensuring users can detect both densely packed areas and isolated outliers. This design was partly inspired by Bubble Cursor~\cite{grossman_bubble_2005} and Bubble Lens~\cite{mott2014beating} that enhance target acquisition, and TactualPlot \cite{chundury2023tactualplot} that enables scalable data exploration using data-to-sound representation, but we employed a different algorithm. Our algorithm for the ``dynamic scanning radius'' feature (\autoref{fig:dtm}e) is provided below in Appendix (Algorithm \ref{alg}).

In our participatory design sessions, we identified a significant challenge for blind users navigating 2-D visualizations like scatter plots: maintaining a consistent movement direction without visual cues. For example, when attempting to explore data points along a specific axis, users often inadvertently shift their movement vertically or horizontally, leading to misinterpretations of a chart's spatial distribution. To counteract this, we implemented a ``directional lock feedback'' mechanism (\autoref{fig:dtm}f) designed to support blind users in accurately navigating the visualization space without increasing cognitive burden. This feature activates when a user moves sequentially in a straight line across three data cells, at which point ChartA11y locks onto that movement direction. Deviations from this locked direction trigger an audio cue---a ``step-up'' or ``step-down'' tone using two musical notes~\cite{zhang_a11yboard_2023}---indicating the user has moved off course. For instance, if a user intended to move horizontally across data cells but accidentally moves to a cell higher up, a step-up tone alerts them to the unintended vertical movement. The user can then adjust their movement downward until a step-down tone indicates a return to the correct horizontal path, ensuring precise and oriented navigation through the chart. 

Additionally, the split-tap for positional information and series switching via swiping (\autoref{fig:dtm}(c)-(d)) remains functional in scatter plots, maintaining consistency across interaction modes.

\subsubsection{Seamless Transition Between DTM and SNF}
Transitioning between modes in screen reader use often poses challenges, particularly in maintaining context, due to the linear focus inherent to screen readers. To mitigate this issue, ChartA11y incorporates a transition mechanism between the DTM and SNF modes. This integration allows users to employ a simple one-finger gesture to toggle between modes while maintaining their current position within the visualization, ensuring continuity and context are preserved. For instance, consider a scenario where a blind user is analyzing a line chart of COVID-19 case trends in Jan 2022 using SNF, which organizes the data into 31 accessible points aligned from left to right. Should the user opt for a more direct interaction through DTM for continuous sonification access, ChartA11y responds by dynamically zooming the chart so that the current chart's $X$-axis is ranged from Jan 1st to 31st. This transition not only preserves the user's location within the data but also enhances their engagement with the visualization by maintaining the integrity of their navigational path.

\section{Co-design with Two Blind Partners}
\label{sec:journey}

To provide the important details of how we reached the final design iteration of ChartA11y, we describe the journey of designing all the different multimodal interactions and multisensory data representations for ChartA11y, especially the challenges we observed when our design partners used touch to interact with charts, and how we iteratively made design decisions for specific features. After introducing the background of our design partners, we describe the evolution of the design in three respects: (1) the Semantic Navigation Framework, (2) the Direct Touch Mapping mode, and (3) how we expanded to scatter plots. Finally, we present how our partners evolved in using ChartA11y to gain insights from charts, which was an impossible task for them prior to this work. 

\subsection{Design Partners}

Our two design partners and paper co-authors (referred to as AS and MA) who are blind played an integral role in our design process. AS, a 31-year-old female, and MA, a 47-year-old male, are professionals at a large tech corporation. Both reported to be totally blind and were recruited through prior connections, committing to a long-term collaboration on the project as active contributors. AS has a background in AI-related fields and experience with charts. She expressed a strong interest and desire in interpreting charts, such as machine learning loss curves, which she previously could not analyze independently and had to rely on sighted colleagues for verbal descriptions. MA also frequently encounters charts in his daily work involving trading, risk management, and fixed income. He has expressed a need to access his personal health data, which is automatically rendered in visual formats by mobile apps, to better understand his health trends.

\subsection{Touch Experience of Charts}

The initiation of our co-design process commenced with a baseline version of ChartA11y. This initial version enabled blind users to experience a simple, single-series line chart from their laptop browser, which was also opened in the ChartA11y mobile application. Initially, due to technical constraints, auditory feedback was provided through the laptop's browser, facilitated by an accessible visualization engine hosted on the server. This configuration aimed to acquaint our blind design partners with the foundational touch experience for chart navigation. In this baseline version, we implemented a simple ``direct touch mapping'' mode, designed to project users' touch inputs onto the $X$-axis and generate corresponding audio tones that mirrored the data trend; no additional interactions described in the above section were implemented yet. Although this touch experience could, in theory, be replicated by the desktop version, Chart Reader, using simple left and right arrow key navigation, leveraging this initial touch-based version of ChartA11y in our co-design sessions was pivotal. It enabled us to identify and address the distinct challenges faced by blind users when interacting with data visualizations via touch. Prior research \cite{zhao_tada_2024, zhang_a11yboard_2023} and feedback from our partners concurred on the unique value that touch interaction brings to the comprehension of visual content. 

Below, we describe how our design partners typically experience charts through touch, uncovering benefits of using touch to access charts and challenges unidentified in prior literature.

\begin{figure*}
    \centering
    \includegraphics[width=\textwidth]{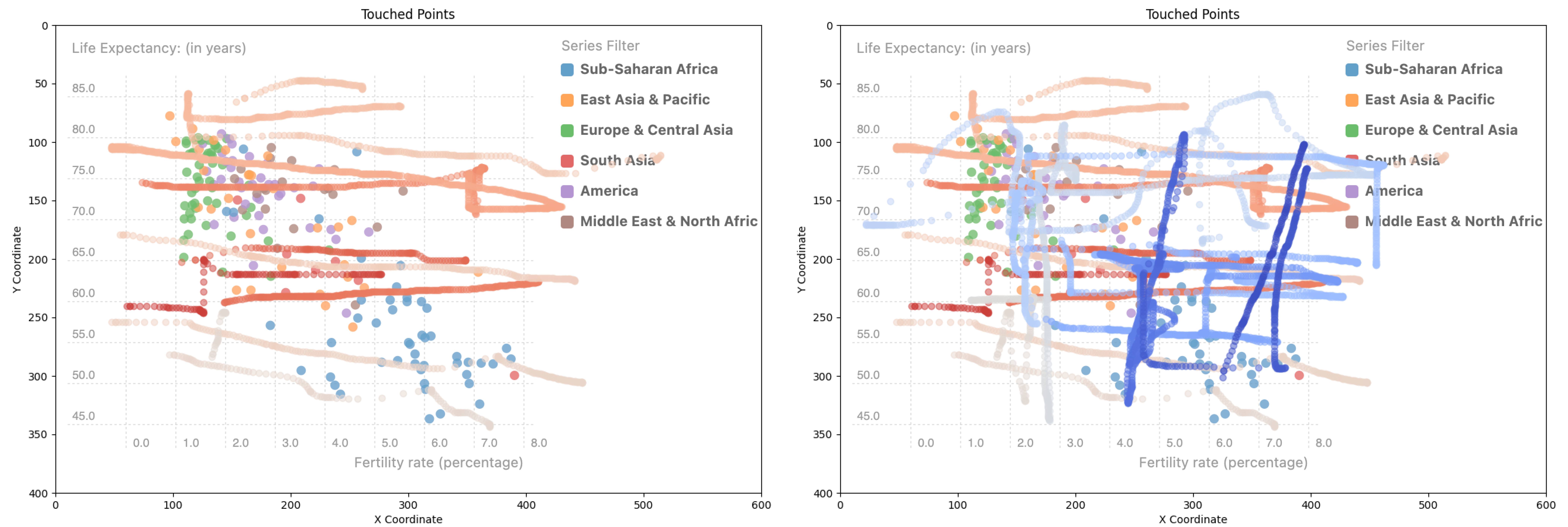}
    \caption{AS's interactions with a scatter plot. The left plot shows the first two minutes of a five-minute interaction, while the right side shows movements from the full five-minute interaction. Colors represent the temporal order of movements: red represents initial movements and blue represents later movements. The same applies for Figure \ref{fig:tp1}.}
    \Description{AS's interactions with a scatter plot. The left side plot is showing the first 2 minutes of a 5-minute interaction time range, while the right side showing movements from a full 5-minute interaction}
    \label{fig:tp2}
\end{figure*}

\subsubsection{Benefits of Touch Experience}
In accessible visualization research, traditional methodologies and existing screen readers have predominantly relied on keyboard interactions. Although efficient for conveying insights authored in charts, keyboard interactions inherently lack direct engagement with digital content. Through our co-design process, partners with experiences in navigating accessible charts via keyboards (e.g., using Chart Reader~\cite{thompson_chart_2023}) identified distinct advantages afforded by the touch experience. 

\paragraph{Spatial Orientation} The touch experience inherently offers a more intuitive spatial awareness by allowing users to physically interact with the device \cite{kane2011usable}. This direct engagement contrasts sharply with keyboard interactions, where the sense of location within a visualization is indirect and intangible. Our design partners highlighted the significant advantage of physically holding and interacting with a device, supporting a preliminary, yet valuable, sense of position on the screen, an aspect notably absent in keyboard-based navigation. A common scenario in the co-design sessions was that both MA and AS utilized their awareness of how close they were to the edges of the screen to locate the desired location~\cite{froehlich2007barrier, wobbrock2003benefits}.

\paragraph{Efficient Navigation} Beyond the initial spatial orientation, the touch interface facilitates a more efficient mechanism for navigating to and, most importantly, re-locating desired data items. Leveraging their spatial memory, our design partners demonstrated the capability to swiftly return to previous positions on the screen after disengaging, a task that was previously inaccessible with keyboard-based screen readers once the focus is lost. For example, when AS was exploring COVID case numbers for August 18th, 2022, she could swiftly return to the specific data item after momentarily lifting her finger from the screen. She landed on August 17th, 2022, and easily swiped right to get the previous location. This efficiency in ``quickly getting back'' to a previously focused element highlights a critical improvement over more basic interactions. Although this feature could be achieved by engineering efforts in keyboard-based screen readers, ChartA11y emphasizes leveraging the innate capabilities of blind users themselves.

\paragraph{Tactile Feedback} Another significant benefit is the enhancement of data visualization tactility through vibrotactile feedback. Unlike auditory feedback, which may introduce delays, the immediacy of localized vibrotactile feedback provides for a more responsive interaction. This real-time feedback, particularly beneficial during rapid navigation, makes visual elements within charts feel more tangible, offering an additional layer of engagement with the data.

\subsubsection{Challenges Occurred}
Below, we summarize the main challenges blind users faced when interacting with charts using touch on smartphones.

\paragraph{Difficulty in Following Visual Traces} A primary challenge we observed was when our partners attempted to trace a variety of visual elements, such as the trend in a line chart, the boundary of scatter plot data clusters, or maintaining a consistent horizontal or vertical direction. Contrary to our initial assumption that users could track trends via changes in audio tones or maintain straight lines by leveraging their awareness of the device's edges, both MA and AS struggled with these tasks. For instance, attempts to scan a scatter plot horizontally (\autoref{fig:tp2}) often resulted in sloped movements, leading to potential misinterpretations of the chart's data correlations. This issue highlighted the difficulty in accurately following visual paths without visual feedback, impacting the users' ability to grasp the intended insights from the data visualization.

\paragraph{Balancing Complex Configuration Options} Just like any other interaction techniques, touch on chart visualizations requires careful consideration, which includes adjusting settings such as switching between data series, changing audio feedback modes, and potentially modifying verbosity levels or number reporting styles. However, the challenge is that the interaction design that effectively maps these configuration needs to the available input mechanisms on a touch screen. The possible input sources include the use of accessibility rotors, custom actions that can be added to any accessibility element, voice commands, unique gesture inputs, and more. It is vital but challenging to design and configure user inputs so that blind smartphone users feel comfortable and confident using them without increasing cognitive burden.

\paragraph{Adapting to Direct Touch Over VoiceOver Gestures} Another substantial challenge emerged from our partners' transition from familiar VoiceOver navigation gestures to the direct touch interactions. Blind smartphone users are typically adept at using specific gestures, like swipes and double taps, to navigate digital content~\cite{kane2011usable}. However, direct touch interactions---essential for engaging with charts in ChartA11y---present a different paradigm. Our partners had some familiarity with direct touch from daily tasks, such as locating and opening apps on the iOS Home Screen. The experience involves remembering the approximate screen location of an app and using touch to activate it directly, a skill that relies on spatial memory and direct touch interaction. Despite this background, applying similar direct touch techniques to the more complex and abstract environment of data visualization was challenging. Navigating arbitrary charts with varied layouts and elements requires a nuanced understanding and adaptation of touch interactions. For example, AS struggled to perform gestures like the split-tap~\cite{kane_slide_2008}, a gesture that is supported in VoiceOver but was not familiar to her. She also had trouble panning on the screen to scan the chart's data area during the first few sessions. The trouble was because what she was familiar with were swiping and double taps, which were iquick to perform, while panning requires continuous focus. Often, she was not aware of how far or close her finger was to the screen before every panning attempt, and had to timidly reach out her finger until it hit the screen.

\paragraph{Orientation Within the Abstract Chart} Although touch experience on a handy device has the potential to give blind users the awareness of their finger's relative position, a persistent challenge was their ability to understand their location within the chart's abstract space. Given the variable layouts and data distributions of different charts, our partners often found themselves lost, struggling to determine their semantic position. 

\subsection{Exploring a Two-Mode Solution}
The benefits and challenges of direct touch that emerged from the co-design sessions showed that both keyboard interactions and simple touch-to-audio interactions have inherent limitations when used alone. Therefore, we developed a two-mode solution, as described in Section \ref{sec:interaction}, including the Semantic Navigation Framework and the Direct Touch Mapping mode. Our goal was that through iteration, these two modes could co-exist, allowing easy transition to one another and providing blind users with different functionality so that they could integrate the knowledge they gained from two modes for a combined understanding of the chart. In this section, we discuss the development of the two-mode solution, showing how the final artifact was driven by findings in the co-design sessions.

We divide the two-mode solution into two major categories, input and output configuration. Output can be further divided into the speech, sonification, and vibration, which are used in SNF and DTM separately.

\subsubsection{Input Configuration}
As we identified the challenge of balancing multiple input sources including but not limited to accessibility rotors, custom actions for accessibility elements in iOS, voice commands, and custom gestures, we implemented several input options for our blind partners to try out and evaluate. For instance, we incorporated the function of multi-series switching into both the rotors and custom actions for assessment. Note that custom actions are defined as a series of supplementary actions that can be appended to an accessibility element in iOS; users can navigate through them by swiping and then double-tap to select. Our partners found that while assigning each series to a custom action was intuitive—owing to the ease of swiping through series at any point—this method was not ideal. In practice, to discern data trends, they commonly toggled between two series for comparison, relying on the auditory cues provided. Configuring series toggling within custom actions required an additional double-tap to switch, which complicated the comparative analysis of data. Conversely, placing series toggling within the accessibility rotor was not only equally accessible but also afforded the added benefit of more direct execution.

Similar nuanced considerations were applied to the design of gestures for functions such as rapid navigation jumps to a specified chart area. In the context of SNF, where VoiceOver remains operative over all elements, gestures are already reserved by VoiceOver and correlated with functions familiar to blind smartphone users. Consequently, assigning an operation to an arbitrary gesture would not be user-friendly, as it would increase the cognitive burden by requiring users to memorize each gesture. We standardized all rapid navigation gestures to start with a double-tap and hold. This action, followed by directional movement (up/down/left/right) or a sustained hold, informs the ChartA11y system of the user's intent to leap to a corresponding area—the X-axis, Y-axis, data point, or filter area—or to transition from the SNF to DTM mode. Both design partners believed that this unified initiation gesture, coupled with semantic linkage to various areas, alleviates the burden of memorizing disparate gestures.

\subsubsection{Output Configuration}
The design of how blind users receive feedback in speech and sonification is another vital aspect of ChartA11y. Echoing the iterative approach applied to input configuration, we deployed a range of intermediate features for evaluation by our design partners. For instance, during their engagement with the SNF on a line chart—where X values served as indices and Y values were the primary focus—it became apparent that the predetermined sequence of speech output could be substantially refined. In the initial design, the accessibility engine generated labels for each data bin or point in a fixed order: X value followed by Y value, and finally, series information. This sequence proved to be problematic when AS navigated across the X-axis. As she moved sequentially from one data item to the next, she found it redundant to hear the X value first, since she was already aware of her relative position. This forced her to pay undue attention with each transition. Yet, when accessing a new and disparate position on the chart, she preferred receiving the positional index information before the Y value. In response to this observation, we realized the need for an adaptive and location-aware approach to speech output that accounts for the user's navigational pattern. We incorporated a mechanism allowing ChartA11y to discern whether a user is moving to an adjacent item or jumping to a remote one, and then decide the sequence in which the positional index or the dependent values should be vocalized first. Similar to the input configuration, such detailed but important iteration was vital to achieve an accessible touch experience on charts.

\subsection{Expanding to Scatter Plots}
Scatter plots, which feature numerous data points distributed across a 2-D space, are inherently more complex than line and bar charts and require even more considerations into how the rich 2-D information can be made accessible without introducing extra burdens.

\begin{figure}
    \centering
    \includegraphics[width=\linewidth]{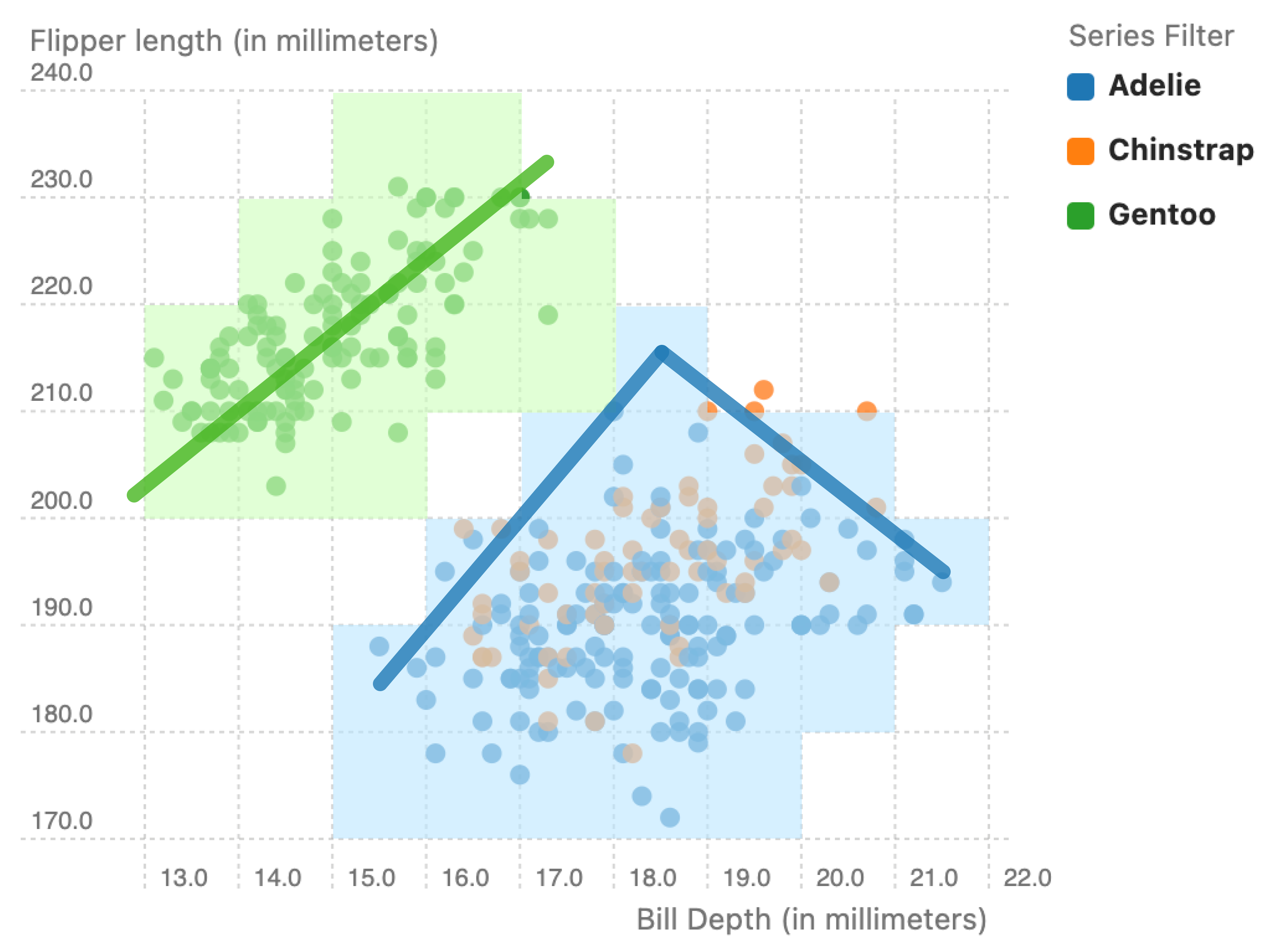}
    \caption{A scatter plot used in co-design, highlighting sonifications by data cells. There is a correlation in the green series, but the aggregated cells made it hard to discern, while the boundary of the blue series was easier for our partners to recognize.}
    \Description{A scatter plot used in co-design sessions, highlighting sonifications aggregated by data cells, drawn in aggregated rectangles. Each rectangle represents a distinct audio tone. There is an obvious correlation in the green Gentoo penguin species, but the aggregated data cell sonification made it hard to identify the correlation. In the meantime, the boundary of the blue Adelie species was easier for our partners to recognize.}
    \label{fig:penguin}
\end{figure}

\begin{figure*}
    \centering
    \includegraphics[width=0.9\textwidth]{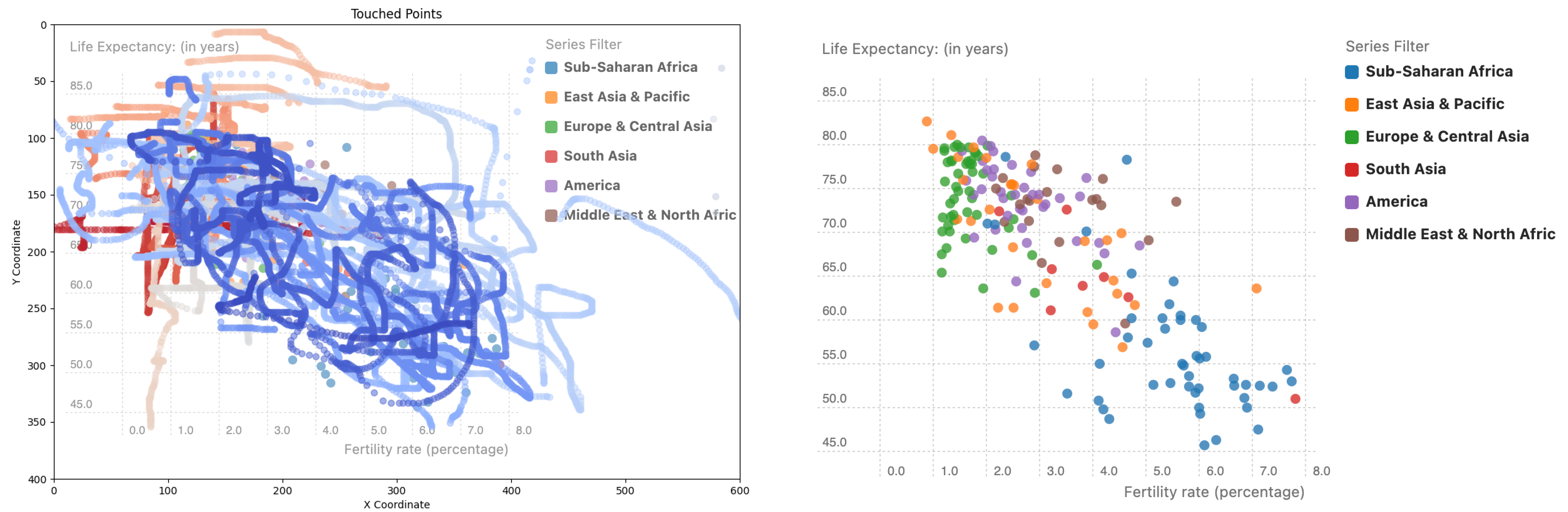}
    \caption{MA's interactions with a scatter plot for 15 minutes, with finger panning over most of the scattered data points. For visibility, we also show the original chart on the right side. The movements gradually converged (deep blue traces) to the data point area upon MA's continuous exploration and understanding of the chart.}
    \Description{MA's interactions with a scatter plot for 15 minutes, with finger panning over most of the scattered data points. The main takeaway in this figure is that the movement finally converged to the data point area range upon users' continuous exploration.}
    \label{fig:tp1}
\end{figure*}

In the context of scatter plots, the foundational approach of employing a dual-mode solution still proved to be useful, as each mode offered distinct advantages for blind smartphone users in comprehending the chart. This became evident during our exploration of a scatter plot, shown in \autoref{fig:penguin}, which rendered data on three penguin species from the Palmer Archipelago (Antarctica) dataset, categorized by bill depth on the X-axis and flipper length on the Y-axis. As our partners engaged with the chart in the SNF mode, they were presented with audio cues representing data points within the vertical bins of the X-axis. For instance, the audio sequence ``tick, tick, tick, beep, beep, tick, tick'' indicated the presence of data points in only the fourth and fifth cells of the Gentoo series' first bin. This enabled them to infer general distribution patterns, leading to accurate conclusions such as the Gentoo species' presence in the upper left quadrant and the overlapping distribution of the other species in the lower right. MA, in particular, could also interpret the cluster shapes, like the `rooftop' form of the Adelie species (blue lines in \autoref{fig:penguin}). However, the aggregated sonifications on data cell only provided a broad view of data distribution. A more granular investigation was necessary to uncover relationships such as the positive correlation between bill depth and flipper length within the Gentoo species, which remained hidden due to the generalized sonification. In such case, the precision of the DTM mode became more useful, giving users refined access when they hit individual data points and revealing details like correlation and data density. Both partners could successfully identify the correlation within the Gentoo series and the different data densities presented in other charts (e.g., \autoref{fig:tp1}).

The synergistic application of the SNF and DTM modes in scatter plots is thus not only complementary but also integral, enabling blind users to derive a comprehensive understanding of such complex visualization.

\subsection{User Experience Example: MA Accessing a Multi-Series Scatter Plot}

To demonstrate how ChartA11y is utilized from a user's perspective, we present an instance where MA engaged with a complex multi-series scatter plot. The scatter plot depicted a dataset's two attributes: life expectancy on the Y-axis and fertility rate on the X-axis (\autoref{fig:tp1}-right). Each data point represents the data of a country, with points aggregated by six global regions. There is a noticeable trend where regions with higher fertility rates often have lower life expectancies, and vice versa, forming a descending pattern from the top left to the bottom right across the plot.

MA started by listening to the descriptions of the chart, including the axis, and data points divided in different series. He then double tapped on the overview description element to get to the main semantic view of X, Y, data point, and series areas (\autoref{fig:snf}). Second, MA chose to access the chart starting from the X-axis area with the SNF mode, and reached his finger to the lower middle half of the screen until he heard ``X axis area.'' He then double tapped again and entered into the X-axis area with nine data bins listed from left to right. Third, MA performed the rotation gesture to switch between all available rotors until he reached ``Sonification,'' and swiped down once. ChartA11y then gave feedback through VoiceOver to confirm that he had turned sonification mode on. He then panned his finger over the screen, and heard six continuous audio tones in sequence for each data bin, indicating if there are data points for each region in this specific data bin. After listening to a few data bin's sonifications, he realized that there are too many series in this chart, so sonification of all six series does not make sense to him. Then, he performed rotation gesture again to reach the ``Series'' rotor, where he could swipe up and down to switch between the series. He then swiped right multiple times to hear sonifications of each cell in each data bin. Through multiple similar trials in different series of regions, MA confirmed that each region has different data distribution. For example, the Sub-Saharan Africa series was mostly empty in the first few data bins, and appeared to be at the first few data cells when reaching the data bins to the right. MA was able to understand that it means the Sub-Saharan Africa data points are generally in the bottom right corner. 

However, despite that MA was able to identify and differentiate rough data distributions between series, which matched his guess, he was only getting sonifications that were aggregated by data cell. The details within the cells were not revealed. For example, he could not identify whether there was any correlation between the X and Y values. He then switched to the DTM mode via the navigational structure, and started the touch-based exploration, whose finger movement was visualized in \autoref{fig:tp1}-left. The exploration lasted for 15 minutes, consisting of interacting with other researchers, asking questions related to usability, and free exploration. The finger movements close to the color red indicate movements that were earlier, while the movements close to blue indicate later movements. MA did a full investigation on the ``overview'' series which displayed all data points and also by each series to compare between them. After exploring, MA was able to identify the clear negative correlation between life expectancy and fertility rate. He was also able to further make spatial sense of each data cluster divided by region. For instance, he noted that Europe \& Central Asia typically exhibited higher life expectancies and lower fertility rates, whereas the Sub-Saharan Africa region had a markedly different data distribution. By encountering fewer data points from the Sub-Saharan African region in the top left corner, he also recognized that the regional distributions were not entirely segregated, with overlaps occurring among a minority of countries.

After the experience, MA could independently access and understand insights in a scatter plot. He described the overall data distribution, correlation among different values, differences between data series, and even the potential existence of outliers. In the meantime, all sighted researchers agreed that what he described was accurate. %
Therefore, MA was able to give conclusions that were close to what sighted people would come up with a glance at the chart, which was not possible prior to using ChartA11y.

\section{Discussion}

We have described ChartAlly, an accessible visualization system powered by touch experience and elaborated on the co-design methodology that led to its creation.
In this section, we further reflect on the design process and the final artifact by discussing the lessons we learned that may be valuable to the community, and discuss the limitations of ChartAlly and directions for future work.

\subsection{Interplay of Modes and Cognitive Load Mitigation}
An important guiding principle consistently observed in our co-design sessions was the goal to enable blind users to navigate complex data visualizations with minimal cognitive strain. We took numerous iterations, often to make what seemed like minor adjustments, but these were in pursuit of significantly reducing cognitive load. For instance, the implementation of a location-aware and adaptive speech output—altering the sequence of X and Y value reporting based on user navigation—may appear as a slight enhancement. Nonetheless, it empowered our partners to proficiently interpret and understand complex charts. 

Another salient finding pertains to the dynamic interaction between the two modes within ChartA11y. It was evident that neither mode alone was sufficient for users to grasp a holistic understanding of the data presented. Only through their combined utility could our partners achieve an integrated comprehension of the charts. We emphasize that the objective of designing assistive technologies is not merely to transcribe all essential data into screen reader announcements or audio synthesis. Achieving theoretical accessibility is a critical step, yet it is the practical usability that truly enhances the experience for screen reader users.

\subsection{Customization is Needed, Not Only in Speech Output}
The principle of customization has been widely recognized as essential, given the diverse needs of users, particularly in making complex visuals like charts accessible. Previous research has explored customization in aspects such as speech output—verbosity, style, sequence, and so on \cite{jones_customization_2024}. We posit that a broader spectrum of customization options is necessary in accessible visualizations, extending to input configurations, sonification tones, haptic feedback patterns, and even the sizes of data bins or cells. Due to the focus of our research scope, we did not exhaustively examine all potential customizations related to sonification techniques and data segmentation. Nevertheless, our observations indicated a clear preference among our design partners for a more adaptable ChartA11y system. The manner in which data is aggregated and sonified can greatly impact the user's perception of the data.

\subsection{User Empowerment in the Age of AI}
In the initial co-design sessions, our partners questioned the necessity of a touch-based interface given the current advancements in AI, positing, ``Why don’t you just throw the chart to ChatGPT and let it tell me what happens in the chart?'' Indeed, recent AI and large language models (LLMs) are capable of analyzing charts to identify trends, correlations, and outliers. Yet, we argue that facilitating users in actively and independently extracting data insights is crucial for fostering inclusion and equity. The process of discovering insights in data is rooted in posing relevant queries and, crucially, understanding which questions to ask. Such discernment comes from direct interaction with the data, a level of engagement not attainable through reliance on generative AI. Furthermore, generative AI is known to hallucinate results, and it becomes even more challenging and vital for accessibility designers and researchers to empower blind users \cite{ladner_design_2015} to verify, comprehend, and refine the AI-generated interpretations in future works.

\subsection{Limitations and Future Work}
In concluding our discussion, we reflect ChartA11y's limitations and outline avenues for future research and development. A notable constraint of our project was in the evaluation methodology. Throughout the co-design sessions, we worked to ensure that our partners engaged with new features and charts using tasks to complete on their own, without sighted researchers' assistance or hints. It allowed us to record accurate reactions and performance from our partners, which would potentially prevent the design from being overly tailored to the preferences of a limited group of stakeholders. Our goal was to ensure that our findings would be broadly applicable to a wide spectrum of blind smartphone users who require data access and interpretation in their daily lives. Despite these efforts, we did not perform a formal user study or deploy ChartA11y to the public. This represents a significant area for future investigation. 

Nonetheless, the extensive design process, enriched by months of collaborative insights and iterations, suggests that ChartA11y transcends a mere prototype. It also has the potential for integration into commercial software, aiming to enhance the accessibility of visualizations on mobile devices. Furthermore, our longitudinal co-design study enabled us to discover previously undocumented challenges of blind users interacting with touch-to-audio methods in a deeper way, which is not possible in a short usability study.

Future enhancements will also focus on integrating greater customization capabilities within the system’s input and output configurations, which is also a current limitation in the system. For example, rotor gestures might be challenging for a number of users since it requires fine finger movement, providing alternatives would be a necessary customization. Finally, exploring the integration of AI to advance the system’s functionality and investigating accessible methods for validating AI-generated results is an obvious and critical area for future work.

\section{Conclusion}

Despite the extensive research on accessible visualizations for blind and low-vision users, little prior work has focused on bringing accessible visualizations from laptops and PCs to smartphones, a place with the potential to provide more direct access to the visual elements. This paper reflects on the iterative co-design process of ChartA11y, which uses multisensory data representations and multimodal interactions to move beyond conventional touch-to-audio experiences. This work also explored accessible scatter plots, a subject that has received minimal attention in prior research. With ChartA11y, our design partners were able to comprehensively understand a complex chart independently, a task that was previously unachievable for them.

\section{Acknowledgement}
This work was supported in part by the Institute of Information and Communications Technology Planning and Evaluation (IITP) Grant funded by the Korean Government (MSIT), Artificial Intelligence Graduate School Program, Yonsei University, under Grant RS-2020-II201361.



\begin{thebibliography}{76}


\ifx \showCODEN    \undefined \def \showCODEN     #1{\unskip}     \fi
\ifx \showDOI      \undefined \def \showDOI       #1{#1}\fi
\ifx \showISBNx    \undefined \def \showISBNx     #1{\unskip}     \fi
\ifx \showISBNxiii \undefined \def \showISBNxiii  #1{\unskip}     \fi
\ifx \showISSN     \undefined \def \showISSN      #1{\unskip}     \fi
\ifx \showLCCN     \undefined \def \showLCCN      #1{\unskip}     \fi
\ifx \shownote     \undefined \def \shownote      #1{#1}          \fi
\ifx \showarticletitle \undefined \def \showarticletitle #1{#1}   \fi
\ifx \showURL      \undefined \def \showURL       {\relax}        \fi
\providecommand\bibfield[2]{#2}
\providecommand\bibinfo[2]{#2}
\providecommand\natexlab[1]{#1}
\providecommand\showeprint[2][]{arXiv:#2}

\bibitem[Access(4 05)]%
        {noauthor_nv_nodate}
\bibfield{author}{\bibinfo{person}{NV Access}.}
  \bibinfo{year}{2024-04-05}\natexlab{}.
\newblock \bibinfo{title}{NVDA}.
\newblock
\newblock
\urldef\tempurl%
\url{https://www.nvaccess.org/}
\showURL{%
\tempurl}


\bibitem[Alam et~al\mbox{.}(2023)]%
        {alam_seechart_2023}
\bibfield{author}{\bibinfo{person}{Md~Zubair~Ibne Alam},
  \bibinfo{person}{Shehnaz Islam}, {and} \bibinfo{person}{Enamul Hoque}.}
  \bibinfo{year}{2023}\natexlab{}.
\newblock \showarticletitle{{SeeChart}: {Enabling} {Accessible}
  {Visualizations} {Through} {Interactive} {Natural} {Language} {Interface}
  {For} {People} with {Visual} {Impairments}}. In
  \bibinfo{booktitle}{\emph{Proceedings of the {International} {Conference} on
  {Intelligent} {User} {Interfaces}}}. \bibinfo{pages}{46--64}.
\newblock
\urldef\tempurl%
\url{https://doi.org/10.1145/3581641.3584099}
\showDOI{\tempurl}
\newblock
\shownote{arXiv:2302.07742 [cs]}.


\bibitem[Awada et~al\mbox{.}(2012)]%
        {awada_towards_2012}
\bibfield{author}{\bibinfo{person}{Amine Awada}, \bibinfo{person}{Youssef~Bou
  Issa}, \bibinfo{person}{Clara Ghannam}, \bibinfo{person}{Joe Tekli}, {and}
  \bibinfo{person}{Richard Chbeir}.} \bibinfo{year}{2012}\natexlab{}.
\newblock \showarticletitle{Towards {Digital} {Image} {Accessibility} for
  {Blind} {Users} {Via} {Vibrating} {Touch} {Screen}: {A} {Feasibility} {Test}
  {Protocol}}. In \bibinfo{booktitle}{\emph{2012 {Eighth} {International}
  {Conference} on {Signal} {Image} {Technology} and {Internet} {Based}
  {Systems}}}. \bibinfo{publisher}{IEEE}, \bibinfo{pages}{547--554}.
\newblock
\showISBNx{978-1-4673-5152-2 978-0-7695-4911-8}
\urldef\tempurl%
\url{https://doi.org/10.1109/SITIS.2012.85}
\showDOI{\tempurl}


\bibitem[Bandukda and Holloway(2020)]%
        {bandukda_audio_2020}
\bibfield{author}{\bibinfo{person}{Maryam Bandukda} {and}
  \bibinfo{person}{Catherine Holloway}.} \bibinfo{year}{2020}\natexlab{}.
\newblock \showarticletitle{Audio {AR} to support nature connectedness in
  people with visual disabilities}. In \bibinfo{booktitle}{\emph{Adjunct
  {Proceedings} of the {ACM} {International} {Joint} {Conference} on
  {Pervasive} and {Ubiquitous} {Computing} and {Proceedings} of the {ACM}
  {International} {Symposium} on {Wearable} {Computers}}}
  \emph{(\bibinfo{series}{{UbiComp}/{ISWC} '20 {Adjunct}})}.
  \bibinfo{publisher}{ACM}, \bibinfo{address}{New York, NY, USA},
  \bibinfo{pages}{204--207}.
\newblock
\showISBNx{978-1-4503-8076-8}
\urldef\tempurl%
\url{https://doi.org/10.1145/3410530.3414332}
\showDOI{\tempurl}


\bibitem[Battle et~al\mbox{.}(2018)]%
        {battle_beagle_2018}
\bibfield{author}{\bibinfo{person}{Leilani Battle}, \bibinfo{person}{Peitong
  Duan}, \bibinfo{person}{Zachery Miranda}, \bibinfo{person}{Dana Mukusheva},
  \bibinfo{person}{Remco Chang}, {and} \bibinfo{person}{Michael Stonebraker}.}
  \bibinfo{year}{2018}\natexlab{}.
\newblock \showarticletitle{Beagle: {Automated} {Extraction} and
  {Interpretation} of {Visualizations} from the {Web}}. In
  \bibinfo{booktitle}{\emph{Proceedings of the {CHI} {Conference} on {Human}
  {Factors} in {Computing} {Systems}}}. \bibinfo{publisher}{ACM},
  \bibinfo{address}{New York, NY, USA}, \bibinfo{pages}{1--8}.
\newblock
\showISBNx{978-1-4503-5620-6}
\urldef\tempurl%
\url{https://doi.org/10.1145/3173574.3174168}
\showDOI{\tempurl}


\bibitem[Blanch et~al\mbox{.}(2004)]%
        {blanch_semantic_2004}
\bibfield{author}{\bibinfo{person}{Renaud Blanch}, \bibinfo{person}{Yves
  Guiard}, {and} \bibinfo{person}{Michel Beaudouin-Lafon}.}
  \bibinfo{year}{2004}\natexlab{}.
\newblock \showarticletitle{Semantic pointing: improving target acquisition
  with control-display ratio adaptation}. In
  \bibinfo{booktitle}{\emph{Proceedings of the {SIGCHI} {Conference} on {Human}
  {Factors} in {Computing} {Systems}}} \emph{(\bibinfo{series}{{CHI} '04})}.
  \bibinfo{publisher}{ACM}, \bibinfo{address}{New York, NY, USA},
  \bibinfo{pages}{519--526}.
\newblock
\showISBNx{978-1-58113-702-6}
\urldef\tempurl%
\url{https://doi.org/10.1145/985692.985758}
\showDOI{\tempurl}


\bibitem[Blanco et~al\mbox{.}(2022)]%
        {blanco_olli_nodate}
\bibfield{author}{\bibinfo{person}{Matthew Blanco}, \bibinfo{person}{Jonathan
  Zong}, {and} \bibinfo{person}{Arvind Satyanarayan}.}
  \bibinfo{year}{2022}\natexlab{}.
\newblock \showarticletitle{Olli: {An} {Extensible} {Visualization} {Library}
  for {Screen} {Reader} {Accessibility}}.
\newblock  (\bibinfo{year}{2022}).
\newblock


\bibitem[Blenkhorn and Evans(1998)]%
        {blenkhorn_using_1998}
\bibfield{author}{\bibinfo{person}{Paul Blenkhorn} {and}
  \bibinfo{person}{D~Gareth Evans}.} \bibinfo{year}{1998}\natexlab{}.
\newblock \showarticletitle{Using speech and touch to enable blind people to
  access schematic diagrams}.
\newblock \bibinfo{journal}{\emph{Journal of Network and Computer
  Applications}} \bibinfo{volume}{21}, \bibinfo{number}{1}
  (\bibinfo{date}{Jan.} \bibinfo{year}{1998}), \bibinfo{pages}{17--29}.
\newblock
\showISSN{10848045}
\urldef\tempurl%
\url{https://doi.org/10.1006/jnca.1998.0060}
\showDOI{\tempurl}


\bibitem[Brown et~al\mbox{.}(2003)]%
        {brown_design_2003}
\bibfield{author}{\bibinfo{person}{Lorna~M Brown}, \bibinfo{person}{Stephen~A
  Brewster}, \bibinfo{person}{Ramesh Ramloll}, \bibinfo{person}{Mike Burton},
  {and} \bibinfo{person}{Beate Riedel}.} \bibinfo{year}{2003}\natexlab{}.
\newblock \showarticletitle{Design guidelines for audio presentation of graphs
  and tables}. In \bibinfo{booktitle}{\emph{9th International Conference on
  Auditory Display (ICAD)}}. \bibinfo{pages}{284--287}.
\newblock
\urldef\tempurl%
\url{http://eprints.gla.ac.uk/3196/}
\showURL{%
\tempurl}


\bibitem[Butler et~al\mbox{.}(2021)]%
        {butler_technology_2021}
\bibfield{author}{\bibinfo{person}{Matthew Butler}, \bibinfo{person}{Leona~M
  Holloway}, \bibinfo{person}{Samuel Reinders}, \bibinfo{person}{Cagatay
  Goncu}, {and} \bibinfo{person}{Kim Marriott}.}
  \bibinfo{year}{2021}\natexlab{}.
\newblock \showarticletitle{Technology {Developments} in {Touch}-{Based}
  {Accessible} {Graphics}: {A} {Systematic} {Review} of {Research} 2010-2020}.
  In \bibinfo{booktitle}{\emph{Proceedings of the {CHI} {Conference} on {Human}
  {Factors} in {Computing} {Systems}}} \emph{(\bibinfo{series}{{CHI} '21})}.
  \bibinfo{publisher}{ACM}, \bibinfo{address}{New York, NY, USA},
  \bibinfo{pages}{1--15}.
\newblock
\showISBNx{978-1-4503-8096-6}
\urldef\tempurl%
\url{https://doi.org/10.1145/3411764.3445207}
\showDOI{\tempurl}


\bibitem[Choi et~al\mbox{.}(2019)]%
        {choi_visualizing_2019}
\bibfield{author}{\bibinfo{person}{Jinho Choi}, \bibinfo{person}{Sanghun Jung},
  \bibinfo{person}{Deok~Gun Park}, \bibinfo{person}{Jaegul Choo}, {and}
  \bibinfo{person}{Niklas Elmqvist}.} \bibinfo{year}{2019}\natexlab{}.
\newblock \showarticletitle{Visualizing for the {Non}-{Visual}: {Enabling} the
  {Visually} {Impaired} to {Use} {Visualization}}.
\newblock \bibinfo{journal}{\emph{Computer Graphics Forum}}
  \bibinfo{volume}{38}, \bibinfo{number}{3} (\bibinfo{year}{2019}),
  \bibinfo{pages}{249--260}.
\newblock
\showISSN{1467-8659}
\urldef\tempurl%
\url{https://doi.org/10.1111/cgf.13686}
\showDOI{\tempurl}
\newblock
\shownote{\_eprint: https://onlinelibrary.wiley.com/doi/pdf/10.1111/cgf.13686}.


\bibitem[Chundury et~al\mbox{.}(2022)]%
        {chundury_towards_2022}
\bibfield{author}{\bibinfo{person}{Pramod Chundury}, \bibinfo{person}{Biswaksen
  Patnaik}, \bibinfo{person}{Yasmin Reyazuddin}, \bibinfo{person}{Christine
  Tang}, \bibinfo{person}{Jonathan Lazar}, {and} \bibinfo{person}{Niklas
  Elmqvist}.} \bibinfo{year}{2022}\natexlab{}.
\newblock \showarticletitle{Towards {Understanding} {Sensory} {Substitution}
  for {Accessible} {Visualization}: {An} {Interview} {Study}}.
\newblock \bibinfo{journal}{\emph{IEEE Transactions on Visualization and
  Computer Graphics}} \bibinfo{volume}{28}, \bibinfo{number}{1}
  (\bibinfo{date}{Jan.} \bibinfo{year}{2022}), \bibinfo{pages}{1084--1094}.
\newblock
\showISSN{1941-0506}
\urldef\tempurl%
\url{https://doi.org/10.1109/TVCG.2021.3114829}
\showDOI{\tempurl}


\bibitem[Chundury et~al\mbox{.}(2023)]%
        {chundury2023tactualplot}
\bibfield{author}{\bibinfo{person}{Pramod Chundury}, \bibinfo{person}{Yasmin
  Reyazuddin}, \bibinfo{person}{J~Bern Jordan}, \bibinfo{person}{Jonathan
  Lazar}, {and} \bibinfo{person}{Niklas Elmqvist}.}
  \bibinfo{year}{2023}\natexlab{}.
\newblock \showarticletitle{TactualPlot: spatializing data as sound using
  sensory substitution for touchscreen accessibility}.
\newblock \bibinfo{journal}{\emph{IEEE Transactions on Visualization and
  Computer Graphics}} (\bibinfo{year}{2023}).
\newblock


\bibitem[Documentation(6 16)]%
        {noauthor_audio_nodate}
\bibfield{author}{\bibinfo{person}{Apple~Developer Documentation}.}
  \bibinfo{year}{2023-06-16}\natexlab{}.
\newblock \bibinfo{title}{Audio graphs}.
\newblock
\newblock
\urldef\tempurl%
\url{https://developer.apple.com/documentation/accessibility/audio_graphs}
\showURL{%
\tempurl}


\bibitem[Elavsky et~al\mbox{.}(2022)]%
        {elavsky_how_2022}
\bibfield{author}{\bibinfo{person}{Frank Elavsky}, \bibinfo{person}{Cynthia
  Bennett}, {and} \bibinfo{person}{Dominik Moritz}.}
  \bibinfo{year}{2022}\natexlab{}.
\newblock \showarticletitle{How accessible is my visualization? {Evaluating}
  visualization accessibility with {Chartability}}.
\newblock \bibinfo{journal}{\emph{Computer Graphics Forum}}
  \bibinfo{volume}{41}, \bibinfo{number}{3} (\bibinfo{date}{June}
  \bibinfo{year}{2022}), \bibinfo{pages}{57--70}.
\newblock
\showISSN{0167-7055, 1467-8659}
\urldef\tempurl%
\url{https://doi.org/10.1111/cgf.14522}
\showDOI{\tempurl}


\bibitem[Fan et~al\mbox{.}(2023)]%
        {fan_accessibility_2023}
\bibfield{author}{\bibinfo{person}{Danyang Fan}, \bibinfo{person}{Alexa
  Fay~Siu}, \bibinfo{person}{Hrishikesh Rao}, \bibinfo{person}{Gene Sung-Ho
  Kim}, \bibinfo{person}{Xavier Vazquez}, \bibinfo{person}{Lucy Greco},
  \bibinfo{person}{Sile O'Modhrain}, {and} \bibinfo{person}{Sean Follmer}.}
  \bibinfo{year}{2023}\natexlab{}.
\newblock \showarticletitle{The {Accessibility} of {Data} {Visualizations} on
  the {Web} for {Screen} {Reader} {Users}: {Practices} and {Experiences}
  {During} {COVID}-19}.
\newblock \bibinfo{journal}{\emph{ACM Transactions on Accessible Computing}}
  \bibinfo{volume}{16}, \bibinfo{number}{1} (\bibinfo{date}{March}
  \bibinfo{year}{2023}), \bibinfo{pages}{1--29}.
\newblock
\showISSN{1936-7228, 1936-7236}
\urldef\tempurl%
\url{https://doi.org/10.1145/3557899}
\showDOI{\tempurl}


\bibitem[Feng(2016)]%
        {feng_designing_2016}
\bibfield{author}{\bibinfo{person}{Catherine Feng}.}
  \bibinfo{year}{2016}\natexlab{}.
\newblock \showarticletitle{Designing {Wearable} {Mobile} {Device}
  {Controllers} for {Blind} {People}: {A} {Co}-{Design} {Approach}}. In
  \bibinfo{booktitle}{\emph{Proceedings of the 18th {International} {ACM}
  {SIGACCESS} {Conference} on {Computers} and {Accessibility}}}
  \emph{(\bibinfo{series}{{ASSETS} '16})}. \bibinfo{publisher}{ACM},
  \bibinfo{address}{New York, NY, USA}, \bibinfo{pages}{341--342}.
\newblock
\showISBNx{978-1-4503-4124-0}
\urldef\tempurl%
\url{https://doi.org/10.1145/2982142.2982144}
\showDOI{\tempurl}


\bibitem[Ferres et~al\mbox{.}(2013)]%
        {ferres_evaluating_2013}
\bibfield{author}{\bibinfo{person}{Leo Ferres}, \bibinfo{person}{Gitte
  Lindgaard}, \bibinfo{person}{Livia Sumegi}, {and} \bibinfo{person}{Bruce
  Tsuji}.} \bibinfo{year}{2013}\natexlab{}.
\newblock \showarticletitle{Evaluating a {Tool} for {Improving} {Accessibility}
  to {Charts} and {Graphs}}.
\newblock \bibinfo{journal}{\emph{ACM Transactions on Computer-Human
  Interaction}} \bibinfo{volume}{20}, \bibinfo{number}{5} (\bibinfo{date}{Nov.}
  \bibinfo{year}{2013}), \bibinfo{pages}{1--32}.
\newblock
\showISSN{1073-0516, 1557-7325}
\urldef\tempurl%
\url{https://doi.org/10.1145/2533682.2533683}
\showDOI{\tempurl}


\bibitem[Flowers et~al\mbox{.}(1997)]%
        {flowers_cross-modal_1997}
\bibfield{author}{\bibinfo{person}{John~H. Flowers}, \bibinfo{person}{Dion~C.
  Buhman}, {and} \bibinfo{person}{Kimberly~D. Turnage}.}
  \bibinfo{year}{1997}\natexlab{}.
\newblock \showarticletitle{Cross-{Modal} {Equivalence} of {Visual} and
  {Auditory} {Scatterplots} for {Exploring} {Bivariate} {Data} {Samples}}.
\newblock \bibinfo{journal}{\emph{Human Factors}} \bibinfo{volume}{39},
  \bibinfo{number}{3} (\bibinfo{date}{Sept.} \bibinfo{year}{1997}),
  \bibinfo{pages}{341--351}.
\newblock
\showISSN{0018-7208}
\urldef\tempurl%
\url{https://doi.org/10.1518/001872097778827151}
\showDOI{\tempurl}
\newblock
\shownote{Publisher: SAGE Publications Inc}.


\bibitem[Froehlich et~al\mbox{.}(2007)]%
        {froehlich2007barrier}
\bibfield{author}{\bibinfo{person}{Jon Froehlich}, \bibinfo{person}{Jacob~O.
  Wobbrock}, {and} \bibinfo{person}{Shaun~K. Kane}.}
  \bibinfo{year}{2007}\natexlab{}.
\newblock \showarticletitle{Barrier pointing: using physical edges to assist
  target acquisition on mobile device touch screens}. In
  \bibinfo{booktitle}{\emph{Proceedings of the 9th International ACM SIGACCESS
  Conference on Computers and Accessibility}} \emph{(\bibinfo{series}{{ASSETS}
  '07})}. \bibinfo{publisher}{ACM}, \bibinfo{address}{New York, NY, USA},
  \bibinfo{pages}{19–26}.
\newblock
\showISBNx{9781595935731}
\urldef\tempurl%
\url{https://doi.org/10.1145/1296843.1296849}
\showDOI{\tempurl}


\bibitem[Gonçalves et~al\mbox{.}(2020)]%
        {goncalves_playing_2020}
\bibfield{author}{\bibinfo{person}{David Gonçalves}, \bibinfo{person}{André
  Rodrigues}, {and} \bibinfo{person}{Tiago Guerreiro}.}
  \bibinfo{year}{2020}\natexlab{}.
\newblock \showarticletitle{Playing {With} {Others}: {Depicting} {Multiplayer}
  {Gaming} {Experiences} of {People} {With} {Visual} {Impairments}}. In
  \bibinfo{booktitle}{\emph{Proceedings of the 22nd {International} {ACM}
  {SIGACCESS} {Conference} on {Computers} and {Accessibility}}}.
  \bibinfo{publisher}{ACM}, \bibinfo{address}{New York, NY, USA},
  \bibinfo{pages}{1--12}.
\newblock
\showISBNx{978-1-4503-7103-2}
\urldef\tempurl%
\url{https://doi.org/10.1145/3373625.3418304}
\showDOI{\tempurl}


\bibitem[Gorlewicz et~al\mbox{.}(2020)]%
        {gorlewicz_design_2020}
\bibfield{author}{\bibinfo{person}{Jenna~L. Gorlewicz},
  \bibinfo{person}{Jennifer~L. Tennison}, \bibinfo{person}{P.~Merlin Uesbeck},
  \bibinfo{person}{Margaret~E. Richard}, \bibinfo{person}{Hari~P. Palani},
  \bibinfo{person}{Andreas Stefik}, \bibinfo{person}{Derrick~W. Smith}, {and}
  \bibinfo{person}{Nicholas~A. Giudice}.} \bibinfo{year}{2020}\natexlab{}.
\newblock \showarticletitle{Design {Guidelines} and {Recommendations} for
  {Multimodal}, {Touchscreen}-based {Graphics}}.
\newblock \bibinfo{journal}{\emph{ACM Transactions on Accessible Computing}}
  \bibinfo{volume}{13}, \bibinfo{number}{3} (\bibinfo{date}{Sept.}
  \bibinfo{year}{2020}), \bibinfo{pages}{1--30}.
\newblock
\showISSN{1936-7228, 1936-7236}
\urldef\tempurl%
\url{https://doi.org/10.1145/3403933}
\showDOI{\tempurl}


\bibitem[Grossman and Balakrishnan(2005)]%
        {grossman_bubble_2005}
\bibfield{author}{\bibinfo{person}{Tovi Grossman} {and} \bibinfo{person}{Ravin
  Balakrishnan}.} \bibinfo{year}{2005}\natexlab{}.
\newblock \showarticletitle{The bubble cursor: enhancing target acquisition by
  dynamic resizing of the cursor's activation area}. In
  \bibinfo{booktitle}{\emph{Proceedings of the {SIGCHI} {Conference} on {Human}
  {Factors} in {Computing} {Systems}}}. \bibinfo{publisher}{ACM},
  \bibinfo{address}{New York, NY, USA}, \bibinfo{pages}{281--290}.
\newblock
\showISBNx{978-1-58113-998-3}
\urldef\tempurl%
\url{https://doi.org/10.1145/1054972.1055012}
\showDOI{\tempurl}


\bibitem[Highcharts(4 05)]%
        {noauthor_interactive_nodate}
\bibfield{author}{\bibinfo{person}{Highcharts}.}
  \bibinfo{year}{2024-04-05}\natexlab{}.
\newblock \bibinfo{title}{Interactive charting library}.
\newblock
\newblock
\urldef\tempurl%
\url{https://www.highcharts.com/blog/homepage21may/}
\showURL{%
\tempurl}


\bibitem[Holloway et~al\mbox{.}(2022)]%
        {holloway_infosonics_2022}
\bibfield{author}{\bibinfo{person}{Leona~M Holloway}, \bibinfo{person}{Cagatay
  Goncu}, \bibinfo{person}{Alon Ilsar}, \bibinfo{person}{Matthew Butler}, {and}
  \bibinfo{person}{Kim Marriott}.} \bibinfo{year}{2022}\natexlab{}.
\newblock \showarticletitle{Infosonics: {Accessible} {Infographics} for
  {People} who are {Blind} using {Sonification} and {Voice}}. In
  \bibinfo{booktitle}{\emph{{CHI} {Conference} on {Human} {Factors} in
  {Computing} {Systems}}}. \bibinfo{publisher}{ACM}, \bibinfo{address}{New
  York, NY, USA}, \bibinfo{pages}{1--13}.
\newblock
\showISBNx{978-1-4503-9157-3}
\urldef\tempurl%
\url{https://doi.org/10.1145/3491102.3517465}
\showDOI{\tempurl}


\bibitem[Hoque et~al\mbox{.}(2023)]%
        {hoque2023accessible}
\bibfield{author}{\bibinfo{person}{Md~Naimul Hoque}, \bibinfo{person}{Md
  Ehtesham-Ul-Haque}, \bibinfo{person}{Niklas Elmqvist}, {and}
  \bibinfo{person}{Syed~Masum Billah}.} \bibinfo{year}{2023}\natexlab{}.
\newblock \showarticletitle{Accessible Data Representation with Natural Sound}.
  In \bibinfo{booktitle}{\emph{Proceedings of the CHI Conference on Human
  Factors in Computing Systems}} \emph{(\bibinfo{series}{CHI '23})}.
  \bibinfo{publisher}{ACM}, \bibinfo{address}{New York, NY, USA}, Article
  \bibinfo{articleno}{826}, \bibinfo{numpages}{19}~pages.
\newblock
\showISBNx{9781450394215}
\urldef\tempurl%
\url{https://doi.org/10.1145/3544548.3581087}
\showDOI{\tempurl}


\bibitem[Inc.(4 05)]%
        {noauthor_jaws_nodate}
\bibfield{author}{\bibinfo{person}{{Freedom}~{Scientific} Inc.}}
  \bibinfo{year}{2024-04-05}\natexlab{}.
\newblock \bibinfo{title}{{JAWS}®}.
\newblock
\newblock
\urldef\tempurl%
\url{https://www.freedomscientific.com/products/software/jaws/}
\showURL{%
\tempurl}


\bibitem[Jones et~al\mbox{.}(2024)]%
        {jones_customization_2024}
\bibfield{author}{\bibinfo{person}{Shuli Jones},
  \bibinfo{person}{Isabella~Pedraza Pineros}, \bibinfo{person}{Daniel Hajas},
  \bibinfo{person}{Jonathan Zong}, {and} \bibinfo{person}{Arvind
  Satyanarayan}.} \bibinfo{year}{2024}\natexlab{}.
\newblock \showarticletitle{“{Customization} is {Key}”: {Reconfigurable}
  {Content} {Tokens} for {Accessible} {Data} {Visualizations}}.
\newblock  (\bibinfo{year}{2024}).
\newblock


\bibitem[Joyner et~al\mbox{.}(2022)]%
        {joyner_visualization_2022}
\bibfield{author}{\bibinfo{person}{Shakila Cherise~S Joyner},
  \bibinfo{person}{Amalia Riegelhuth}, \bibinfo{person}{Kathleen Garrity},
  \bibinfo{person}{Yea-Seul Kim}, {and} \bibinfo{person}{Nam~Wook Kim}.}
  \bibinfo{year}{2022}\natexlab{}.
\newblock \showarticletitle{Visualization {Accessibility} in the {Wild}:
  {Challenges} {Faced} by {Visualization} {Designers}}. In
  \bibinfo{booktitle}{\emph{{CHI} {Conference} on {Human} {Factors} in
  {Computing} {Systems}}}. \bibinfo{publisher}{ACM}, \bibinfo{address}{New
  York, NY, USA}, \bibinfo{pages}{1--19}.
\newblock
\showISBNx{978-1-4503-9157-3}
\urldef\tempurl%
\url{https://doi.org/10.1145/3491102.3517630}
\showDOI{\tempurl}


\bibitem[Jung et~al\mbox{.}(2022)]%
        {jung_communicating_2022}
\bibfield{author}{\bibinfo{person}{Crescentia Jung}, \bibinfo{person}{Shubham
  Mehta}, \bibinfo{person}{Atharva Kulkarni}, \bibinfo{person}{Yuhang Zhao},
  {and} \bibinfo{person}{Yea-Seul Kim}.} \bibinfo{year}{2022}\natexlab{}.
\newblock \showarticletitle{Communicating {Visualizations} without {Visuals}:
  {Investigation} of {Visualization} {Alternative} {Text} for {People} with
  {Visual} {Impairments}}.
\newblock \bibinfo{journal}{\emph{IEEE Transactions on Visualization and
  Computer Graphics}} \bibinfo{volume}{28}, \bibinfo{number}{1}
  (\bibinfo{date}{Jan.} \bibinfo{year}{2022}), \bibinfo{pages}{1095--1105}.
\newblock
\showISSN{1077-2626, 1941-0506, 2160-9306}
\urldef\tempurl%
\url{https://doi.org/10.1109/TVCG.2021.3114846}
\showDOI{\tempurl}


\bibitem[Jung et~al\mbox{.}(2017)]%
        {jung_chartsense_2017}
\bibfield{author}{\bibinfo{person}{Daekyoung Jung}, \bibinfo{person}{Wonjae
  Kim}, \bibinfo{person}{Hyunjoo Song}, \bibinfo{person}{Jeong-in Hwang},
  \bibinfo{person}{Bongshin Lee}, \bibinfo{person}{Bohyoung Kim}, {and}
  \bibinfo{person}{Jinwook Seo}.} \bibinfo{year}{2017}\natexlab{}.
\newblock \showarticletitle{{ChartSense}: {Interactive} {Data} {Extraction}
  from {Chart} {Images}}. In \bibinfo{booktitle}{\emph{Proceedings of the {CHI}
  {Conference} on {Human} {Factors} in {Computing} {Systems}}}
  \emph{(\bibinfo{series}{{CHI} '17})}. \bibinfo{publisher}{ACM},
  \bibinfo{address}{New York, NY, USA}, \bibinfo{pages}{6706--6717}.
\newblock
\showISBNx{978-1-4503-4655-9}
\urldef\tempurl%
\url{https://doi.org/10.1145/3025453.3025957}
\showDOI{\tempurl}


\bibitem[Kane et~al\mbox{.}(2008)]%
        {kane_slide_2008}
\bibfield{author}{\bibinfo{person}{Shaun~K. Kane}, \bibinfo{person}{Jeffrey~P.
  Bigham}, {and} \bibinfo{person}{Jacob~O. Wobbrock}.}
  \bibinfo{year}{2008}\natexlab{}.
\newblock \showarticletitle{Slide rule: making mobile touch screens accessible
  to blind people using multi-touch interaction techniques}. In
  \bibinfo{booktitle}{\emph{Proceedings of the 10th international {ACM}
  {SIGACCESS} conference on {Computers} and accessibility}}
  \emph{(\bibinfo{series}{{ASSETS} '08})}. \bibinfo{publisher}{ACM},
  \bibinfo{address}{New York, NY, USA}, \bibinfo{pages}{73--80}.
\newblock
\showISBNx{978-1-59593-976-0}
\urldef\tempurl%
\url{https://doi.org/10.1145/1414471.1414487}
\showDOI{\tempurl}


\bibitem[Kane et~al\mbox{.}(2011)]%
        {kane2011usable}
\bibfield{author}{\bibinfo{person}{Shaun~K. Kane}, \bibinfo{person}{Jacob~O.
  Wobbrock}, {and} \bibinfo{person}{Richard~E. Ladner}.}
  \bibinfo{year}{2011}\natexlab{}.
\newblock \showarticletitle{Usable gestures for blind people: understanding
  preference and performance}. In \bibinfo{booktitle}{\emph{Proceedings of the
  CHI Conference on Human Factors in Computing Systems}}
  \emph{(\bibinfo{series}{CHI '11})}. \bibinfo{publisher}{ACM},
  \bibinfo{address}{New York, NY, USA}, \bibinfo{pages}{413–422}.
\newblock
\showISBNx{9781450302289}
\urldef\tempurl%
\url{https://doi.org/10.1145/1978942.1979001}
\showDOI{\tempurl}


\bibitem[Kim and McCoy(2018)]%
        {kim_multimodal_2018}
\bibfield{author}{\bibinfo{person}{Edward Kim} {and}
  \bibinfo{person}{Kathleen~F. McCoy}.} \bibinfo{year}{2018}\natexlab{}.
\newblock \showarticletitle{Multimodal {Deep} {Learning} using {Images} and
  {Text} for {Information} {Graphic} {Classification}}. In
  \bibinfo{booktitle}{\emph{Proceedings of the 20th {International} {ACM}
  {SIGACCESS} {Conference} on {Computers} and {Accessibility}}}
  \emph{(\bibinfo{series}{{ASSETS} '18})}. \bibinfo{publisher}{ACM},
  \bibinfo{address}{New York, NY, USA}, \bibinfo{pages}{143--148}.
\newblock
\showISBNx{978-1-4503-5650-3}
\urldef\tempurl%
\url{https://doi.org/10.1145/3234695.3236357}
\showDOI{\tempurl}


\bibitem[Kim et~al\mbox{.}(2021b)]%
        {kim_information_2021}
\bibfield{author}{\bibinfo{person}{Edward Kim}, \bibinfo{person}{Connor
  Onweller}, {and} \bibinfo{person}{Kathleen~F. McCoy}.}
  \bibinfo{year}{2021}\natexlab{b}.
\newblock \showarticletitle{Information {Graphic} {Summarization} using a
  {Collection} of {Multimodal} {Deep} {Neural} {Networks}}. In
  \bibinfo{booktitle}{\emph{2020 25th {International} {Conference} on {Pattern}
  {Recognition} ({ICPR})}}. \bibinfo{pages}{10188--10195}.
\newblock
\urldef\tempurl%
\url{https://doi.org/10.1109/ICPR48806.2021.9412146}
\showDOI{\tempurl}
\newblock
\shownote{ISSN: 1051-4651}.


\bibitem[Kim et~al\mbox{.}(2021a)]%
        {kim_accessible_2021}
\bibfield{author}{\bibinfo{person}{Nam~Wook Kim},
  \bibinfo{person}{Shakila~Cherise Joyner}, \bibinfo{person}{Amalia
  Riegelhuth}, {and} \bibinfo{person}{Yea-Seul Kim}.}
  \bibinfo{year}{2021}\natexlab{a}.
\newblock \showarticletitle{Accessible {Visualization}: {Design} {Space},
  {Opportunities}, and {Challenges}}.
\newblock \bibinfo{journal}{\emph{Computer Graphics Forum}}
  \bibinfo{volume}{40}, \bibinfo{number}{3} (\bibinfo{date}{June}
  \bibinfo{year}{2021}), \bibinfo{pages}{173--188}.
\newblock
\showISSN{0167-7055, 1467-8659}
\urldef\tempurl%
\url{https://doi.org/10.1111/cgf.14298}
\showDOI{\tempurl}


\bibitem[Ladner(2015)]%
        {ladner_design_2015}
\bibfield{author}{\bibinfo{person}{Richard~E. Ladner}.}
  \bibinfo{year}{2015}\natexlab{}.
\newblock \showarticletitle{Design for user empowerment}.
\newblock \bibinfo{journal}{\emph{Interactions}} \bibinfo{volume}{22},
  \bibinfo{number}{2} (\bibinfo{date}{Feb.} \bibinfo{year}{2015}),
  \bibinfo{pages}{24--29}.
\newblock
\showISSN{1072-5520, 1558-3449}
\urldef\tempurl%
\url{https://doi.org/10.1145/2723869}
\showDOI{\tempurl}


\bibitem[Lazar et~al\mbox{.}(2007)]%
        {lazar_what_2007}
\bibfield{author}{\bibinfo{person}{Jonathan Lazar}, \bibinfo{person}{Aaron
  Allen}, \bibinfo{person}{Jason Kleinman}, {and} \bibinfo{person}{Chris
  Malarkey}.} \bibinfo{year}{2007}\natexlab{}.
\newblock \showarticletitle{What {Frustrates} {Screen} {Reader} {Users} on the
  {Web}: {A} {Study} of 100 {Blind} {Users}}.
\newblock \bibinfo{journal}{\emph{International Journal of Human–Computer
  Interaction}} \bibinfo{volume}{22}, \bibinfo{number}{3} (\bibinfo{date}{May}
  \bibinfo{year}{2007}), \bibinfo{pages}{247--269}.
\newblock
\showISSN{1044-7318}
\urldef\tempurl%
\url{https://doi.org/10.1080/10447310709336964}
\showDOI{\tempurl}


\bibitem[Lee et~al\mbox{.}(2022)]%
        {lee_collabally_2022}
\bibfield{author}{\bibinfo{person}{Cheuk Yin~Phipson Lee},
  \bibinfo{person}{Zhuohao Zhang}, \bibinfo{person}{Jaylin Herskovitz},
  \bibinfo{person}{JooYoung Seo}, {and} \bibinfo{person}{Anhong Guo}.}
  \bibinfo{year}{2022}\natexlab{}.
\newblock \showarticletitle{{CollabAlly}: {Accessible} {Collaboration}
  {Awareness} in {Document} {Editing}}. In \bibinfo{booktitle}{\emph{{CHI}
  {Conference} on {Human} {Factors} in {Computing} {Systems}}}.
  \bibinfo{publisher}{ACM}, \bibinfo{address}{New York, NY, USA},
  \bibinfo{pages}{1--17}.
\newblock
\showISBNx{978-1-4503-9157-3}
\urldef\tempurl%
\url{https://doi.org/10.1145/3491102.3517635}
\showDOI{\tempurl}


\bibitem[Li et~al\mbox{.}(2023)]%
        {li_toucha11y_2023}
\bibfield{author}{\bibinfo{person}{Jiasheng Li}, \bibinfo{person}{Zeyu Yan},
  \bibinfo{person}{Arush Shah}, \bibinfo{person}{Jonathan Lazar}, {and}
  \bibinfo{person}{Huaishu Peng}.} \bibinfo{year}{2023}\natexlab{}.
\newblock \showarticletitle{Toucha11y: {Making} {Inaccessible} {Public}
  {Touchscreens} {Accessible}}. In \bibinfo{booktitle}{\emph{Proceedings of the
  {CHI} {Conference} on {Human} {Factors} in {Computing} {Systems}}}.
  \bibinfo{publisher}{ACM}, \bibinfo{address}{New York, NY, USA},
  \bibinfo{pages}{1--13}.
\newblock
\showISBNx{978-1-4503-9421-5}
\urldef\tempurl%
\url{https://doi.org/10.1145/3544548.3581254}
\showDOI{\tempurl}


\bibitem[Liang et~al\mbox{.}(2023)]%
        {liang_brushlens_2023}
\bibfield{author}{\bibinfo{person}{Chen Liang}, \bibinfo{person}{Yasha
  Iravantchi}, \bibinfo{person}{Thomas Krolikowski}, \bibinfo{person}{Ruijie
  Geng}, \bibinfo{person}{Alanson~P. Sample}, {and} \bibinfo{person}{Anhong
  Guo}.} \bibinfo{year}{2023}\natexlab{}.
\newblock \showarticletitle{{BrushLens}: {Hardware} {Interaction} {Proxies} for
  {Accessible} {Touchscreen} {Interface} {Actuation}}. In
  \bibinfo{booktitle}{\emph{Proceedings of the 36th {Annual} {ACM} {Symposium}
  on {User} {Interface} {Software} and {Technology}}}.
  \bibinfo{publisher}{ACM}, \bibinfo{address}{New York, NY, USA},
  \bibinfo{pages}{1--17}.
\newblock
\showISBNx{9798400701320}
\urldef\tempurl%
\url{https://doi.org/10.1145/3586183.3606730}
\showDOI{\tempurl}


\bibitem[Liu et~al\mbox{.}(2019)]%
        {liu_data_2019}
\bibfield{author}{\bibinfo{person}{Xiaoyi Liu}, \bibinfo{person}{Diego
  Klabjan}, {and} \bibinfo{person}{Patrick NBless}.}
  \bibinfo{year}{2019}\natexlab{}.
\newblock \bibinfo{title}{Data {Extraction} from {Charts} via {Single} {Deep}
  {Neural} {Network}}.
\newblock
\newblock
\urldef\tempurl%
\url{http://arxiv.org/abs/1906.11906}
\showURL{%
\tempurl}


\bibitem[Lundgard et~al\mbox{.}(2019)]%
        {lundgard_sociotechnical_2019}
\bibfield{author}{\bibinfo{person}{Alan Lundgard}, \bibinfo{person}{Crystal
  Lee}, {and} \bibinfo{person}{Arvind Satyanarayan}.}
  \bibinfo{year}{2019}\natexlab{}.
\newblock \showarticletitle{Sociotechnical {Considerations} for {Accessible}
  {Visualization} {Design}}. In \bibinfo{booktitle}{\emph{2019 {IEEE}
  {Visualization} {Conference} ({VIS})}}. \bibinfo{pages}{16--20}.
\newblock
\urldef\tempurl%
\url{https://doi.org/10.1109/VISUAL.2019.8933762}
\showDOI{\tempurl}


\bibitem[Lundgard and Satyanarayan(2022)]%
        {lundgard_accessible_2022}
\bibfield{author}{\bibinfo{person}{Alan Lundgard} {and} \bibinfo{person}{Arvind
  Satyanarayan}.} \bibinfo{year}{2022}\natexlab{}.
\newblock \showarticletitle{Accessible {Visualization} via {Natural} {Language}
  {Descriptions}: {A} {Four}-{Level} {Model} of {Semantic} {Content}}.
\newblock \bibinfo{journal}{\emph{IEEE Transactions on Visualization and
  Computer Graphics}} \bibinfo{volume}{28}, \bibinfo{number}{1}
  (\bibinfo{date}{Jan.} \bibinfo{year}{2022}), \bibinfo{pages}{1073--1083}.
\newblock
\showISSN{1941-0506}
\urldef\tempurl%
\url{https://doi.org/10.1109/TVCG.2021.3114770}
\showDOI{\tempurl}


\bibitem[Ma et~al\mbox{.}(2021)]%
        {ma_towards_2021}
\bibfield{author}{\bibinfo{person}{Weihong Ma}, \bibinfo{person}{Hesuo Zhang},
  \bibinfo{person}{Shuang Yan}, \bibinfo{person}{Guangshun Yao},
  \bibinfo{person}{Yichao Huang}, \bibinfo{person}{Hui Li},
  \bibinfo{person}{Yaqiang Wu}, {and} \bibinfo{person}{Lianwen Jin}.}
  \bibinfo{year}{2021}\natexlab{}.
\newblock \showarticletitle{Towards an Efficient Framework for Data Extraction
  from Chart Images}. In \bibinfo{booktitle}{\emph{Document Analysis and
  Recognition -- ICDAR 2021}}, \bibfield{editor}{\bibinfo{person}{Josep
  Llad{\'o}s}, \bibinfo{person}{Daniel Lopresti}, {and}
  \bibinfo{person}{Seiichi Uchida}} (Eds.). \bibinfo{publisher}{Springer
  International Publishing}, \bibinfo{address}{Cham},
  \bibinfo{pages}{583--597}.
\newblock
\showISBNx{978-3-030-86549-8}


\bibitem[Marriott et~al\mbox{.}(2021)]%
        {marriott_inclusive_2021}
\bibfield{author}{\bibinfo{person}{Kim Marriott}, \bibinfo{person}{Bongshin
  Lee}, \bibinfo{person}{Matthew Butler}, \bibinfo{person}{Ed Cutrell},
  \bibinfo{person}{Kirsten Ellis}, \bibinfo{person}{Cagatay Goncu},
  \bibinfo{person}{Marti Hearst}, \bibinfo{person}{Kathleen McCoy}, {and}
  \bibinfo{person}{Danielle~Albers Szafir}.} \bibinfo{year}{2021}\natexlab{}.
\newblock \showarticletitle{Inclusive data visualization for people with
  disabilities: a call to action}.
\newblock \bibinfo{journal}{\emph{Interactions}} \bibinfo{volume}{28},
  \bibinfo{number}{3} (\bibinfo{date}{May} \bibinfo{year}{2021}),
  \bibinfo{pages}{47--51}.
\newblock
\showISSN{1072-5520, 1558-3449}
\urldef\tempurl%
\url{https://doi.org/10.1145/3457875}
\showDOI{\tempurl}


\bibitem[McGookin and Brewster(2006)]%
        {mcgookin_soundbar_2006}
\bibfield{author}{\bibinfo{person}{David~K. McGookin} {and}
  \bibinfo{person}{Stephen~A. Brewster}.} \bibinfo{year}{2006}\natexlab{}.
\newblock \showarticletitle{{SoundBar}: exploiting multiple views in multimodal
  graph browsing}. In \bibinfo{booktitle}{\emph{Proceedings of the 4th {Nordic}
  conference on {Human}-computer interaction: changing roles}}
  \emph{(\bibinfo{series}{{NordiCHI} '06})}. \bibinfo{publisher}{ACM},
  \bibinfo{address}{New York, NY, USA}, \bibinfo{pages}{145--154}.
\newblock
\showISBNx{978-1-59593-325-6}
\urldef\tempurl%
\url{https://doi.org/10.1145/1182475.1182491}
\showDOI{\tempurl}


\bibitem[Mendes et~al\mbox{.}(2020)]%
        {mendes_collaborative_2020}
\bibfield{author}{\bibinfo{person}{Daniel Mendes}, \bibinfo{person}{Sofia
  Reis}, \bibinfo{person}{João Guerreiro}, {and} \bibinfo{person}{Hugo
  Nicolau}.} \bibinfo{year}{2020}\natexlab{}.
\newblock \showarticletitle{Collaborative {Tabletops} for {Blind} {People}:
  {The} {Effect} of {Auditory} {Design} on {Workspace} {Awareness}}.
\newblock \bibinfo{journal}{\emph{Proceedings of the ACM on Human-Computer
  Interaction}} \bibinfo{volume}{4}, \bibinfo{number}{ISS}
  (\bibinfo{date}{Nov.} \bibinfo{year}{2020}), \bibinfo{pages}{197:1--197:19}.
\newblock
\urldef\tempurl%
\url{https://doi.org/10.1145/3427325}
\showDOI{\tempurl}


\bibitem[{Microsoft Research}(2021)]%
        {microsoft-soundscape}
\bibfield{author}{\bibinfo{person}{{Microsoft Research}}.}
  \bibinfo{year}{2021}\natexlab{}.
\newblock \bibinfo{title}{{Microsoft Soundscape -- A map delivered in 3D
  sound}}.
\newblock
  \bibinfo{howpublished}{\url{https://www.microsoft.com/en-us/research/product/soundscape/}}.
\newblock


\bibitem[Morrison et~al\mbox{.}(2021)]%
        {morrison_social_2021}
\bibfield{author}{\bibinfo{person}{Cecily Morrison}, \bibinfo{person}{Edward
  Cutrell}, \bibinfo{person}{Martin Grayson}, \bibinfo{person}{Anja Thieme},
  \bibinfo{person}{Alex Taylor}, \bibinfo{person}{Geert Roumen},
  \bibinfo{person}{Camilla Longden}, \bibinfo{person}{Sebastian Tschiatschek},
  \bibinfo{person}{Rita Faia~Marques}, {and} \bibinfo{person}{Abigail Sellen}.}
  \bibinfo{year}{2021}\natexlab{}.
\newblock \showarticletitle{Social {Sensemaking} with {AI}: {Designing} an
  {Open}-ended {AI} {Experience} with a {Blind} {Child}}. In
  \bibinfo{booktitle}{\emph{Proceedings of the {CHI} {Conference} on {Human}
  {Factors} in {Computing} {Systems}}} \emph{(\bibinfo{series}{{CHI} '21})}.
  \bibinfo{publisher}{ACM}, \bibinfo{address}{New York, NY, USA},
  \bibinfo{pages}{1--14}.
\newblock
\showISBNx{978-1-4503-8096-6}
\urldef\tempurl%
\url{https://doi.org/10.1145/3411764.3445290}
\showDOI{\tempurl}


\bibitem[Mott and Wobbrock(2014)]%
        {mott2014beating}
\bibfield{author}{\bibinfo{person}{Martez~E. Mott} {and}
  \bibinfo{person}{Jacob~O. Wobbrock}.} \bibinfo{year}{2014}\natexlab{}.
\newblock \showarticletitle{Beating the bubble: using kinematic triggering in
  the bubble lens for acquiring small, dense targets}. In
  \bibinfo{booktitle}{\emph{Proceedings of the CHI Conference on Human Factors
  in Computing Systems}} \emph{(\bibinfo{series}{CHI '14})}.
  \bibinfo{publisher}{ACM}, \bibinfo{address}{New York, NY, USA},
  \bibinfo{pages}{733–742}.
\newblock
\showISBNx{9781450324731}
\urldef\tempurl%
\url{https://doi.org/10.1145/2556288.2557410}
\showDOI{\tempurl}


\bibitem[Mullenbach et~al\mbox{.}(2013)]%
        {mullenbach_surface_2013}
\bibfield{author}{\bibinfo{person}{Joe Mullenbach}, \bibinfo{person}{Craig
  Shultz}, \bibinfo{person}{Anne~Marie Piper}, \bibinfo{person}{Michael
  Peshkin}, {and} \bibinfo{person}{J.~Edward Colgate}.}
  \bibinfo{year}{2013}\natexlab{}.
\newblock \showarticletitle{Surface haptic interactions with a {TPad} tablet}.
  In \bibinfo{booktitle}{\emph{Adjunct Proceedings of the 26th annual {ACM}
  symposium on {User} interface software and technology}}
  \emph{(\bibinfo{series}{{UIST} '13 {Adjunct}})}. \bibinfo{publisher}{ACM},
  \bibinfo{address}{New York, NY, USA}, \bibinfo{pages}{7--8}.
\newblock
\showISBNx{978-1-4503-2406-9}
\urldef\tempurl%
\url{https://doi.org/10.1145/2508468.2514929}
\showDOI{\tempurl}


\bibitem[Obeid and Hoque(2020)]%
        {obeid_chart--text_2020}
\bibfield{author}{\bibinfo{person}{Jason Obeid} {and} \bibinfo{person}{Enamul
  Hoque}.} \bibinfo{year}{2020}\natexlab{}.
\newblock \bibinfo{title}{Chart-to-{Text}: {Generating} {Natural} {Language}
  {Descriptions} for {Charts} by {Adapting} the {Transformer} {Model}}.
\newblock
\newblock
\urldef\tempurl%
\url{http://arxiv.org/abs/2010.09142}
\showURL{%
\tempurl}


\bibitem[Palani et~al\mbox{.}(2018)]%
        {antona_haptic_2018}
\bibfield{author}{\bibinfo{person}{Hari~Prasath Palani},
  \bibinfo{person}{G.~Bernard Giudice}, {and} \bibinfo{person}{Nicholas~A.
  Giudice}.} \bibinfo{year}{2018}\natexlab{}.
\newblock \showarticletitle{Haptic {Information} {Access} {Using} {Touchscreen}
  {Devices}: {Design} {Guidelines} for {Accurate} {Perception} of {Angular}
  {Magnitude} and {Line} {Orientation}}.
\newblock In \bibinfo{booktitle}{\emph{Universal {Access} in {Human}-{Computer}
  {Interaction}. {Methods}, {Technologies}, and {Users}}},
  \bibfield{editor}{\bibinfo{person}{Margherita Antona} {and}
  \bibinfo{person}{Constantine Stephanidis}} (Eds.).
  Vol.~\bibinfo{volume}{10907}. \bibinfo{publisher}{Springer International
  Publishing}, \bibinfo{address}{Cham}, \bibinfo{pages}{243--255}.
\newblock
\showISBNx{978-3-319-92048-1 978-3-319-92049-8}
\urldef\tempurl%
\url{https://doi.org/10.1007/978-3-319-92049-8_18}
\showDOI{\tempurl}
\newblock
\shownote{Series Title: Lecture Notes in Computer Science}.


\bibitem[Robinson~Moore et~al\mbox{.}(2024)]%
        {moore_spatial_2024}
\bibfield{author}{\bibinfo{person}{Wilfredo~Joshua Robinson~Moore},
  \bibinfo{person}{Medhani Kalal}, \bibinfo{person}{Jennifer~L. Tennison},
  \bibinfo{person}{Nicholas~A Giudice}, {and} \bibinfo{person}{Jenna
  Gorlewicz}.} \bibinfo{year}{2024}\natexlab{}.
\newblock \showarticletitle{Spatial Audio-Enhanced Multimodal Graph Rendering
  for Efficient Data Trend Learning on Touchscreen Devices}. In
  \bibinfo{booktitle}{\emph{Proceedings of the CHI Conference on Human Factors
  in Computing Systems}} \emph{(\bibinfo{series}{CHI '24})}.
  \bibinfo{publisher}{ACM}, \bibinfo{address}{New York, NY, USA}, Article
  \bibinfo{articleno}{206}, \bibinfo{numpages}{12}~pages.
\newblock
\showISBNx{9798400703300}
\urldef\tempurl%
\url{https://doi.org/10.1145/3613904.3641959}
\showDOI{\tempurl}


\bibitem[Savva et~al\mbox{.}(2011)]%
        {savva_revision_2011}
\bibfield{author}{\bibinfo{person}{Manolis Savva}, \bibinfo{person}{Nicholas
  Kong}, \bibinfo{person}{Arti Chhajta}, \bibinfo{person}{Li Fei-Fei},
  \bibinfo{person}{Maneesh Agrawala}, {and} \bibinfo{person}{Jeffrey Heer}.}
  \bibinfo{year}{2011}\natexlab{}.
\newblock \showarticletitle{{ReVision}: automated classification, analysis and
  redesign of chart images}. In \bibinfo{booktitle}{\emph{Proceedings of the
  24th annual {ACM} symposium on {User} interface software and technology}}.
  \bibinfo{publisher}{ACM}, \bibinfo{address}{New York, NY, USA},
  \bibinfo{pages}{393--402}.
\newblock
\showISBNx{978-1-4503-0716-1}
\urldef\tempurl%
\url{https://doi.org/10.1145/2047196.2047247}
\showDOI{\tempurl}


\bibitem[Scoy et~al\mbox{.}(2005)]%
        {scoy_auditory_2005}
\bibfield{author}{\bibinfo{person}{Frances~Van Scoy}, \bibinfo{person}{Don
  McLaughlin}, {and} \bibinfo{person}{Angela Fullmer}.}
  \bibinfo{year}{2005}\natexlab{}.
\newblock \showarticletitle{{AUDITORY} {AUGMENTATION} {OF} {HAPTIC} {GRAPHS}:
  {DEVELOPING} {A} {GRAPHIC} {TOOL} {FOR} {TEACHING} {PRECALCULUS} {SKILL} {TO}
  {BLIND} {STUDENTS}}.
\newblock  (\bibinfo{year}{2005}).
\newblock


\bibitem[Seo et~al\mbox{.}(2024)]%
        {seo_maidr_2024}
\bibfield{author}{\bibinfo{person}{JooYoung Seo}, \bibinfo{person}{Yilin Xia},
  \bibinfo{person}{Bongshin Lee}, \bibinfo{person}{Sean Mccurry}, {and}
  \bibinfo{person}{Yu~Jun Yam}.} \bibinfo{year}{2024}\natexlab{}.
\newblock \showarticletitle{MAIDR: Making Statistical Visualizations Accessible
  with Multimodal Data Representation}. In
  \bibinfo{booktitle}{\emph{Proceedings of the CHI Conference on Human Factors
  in Computing Systems}} \emph{(\bibinfo{series}{CHI '24})}.
  \bibinfo{publisher}{ACM}, \bibinfo{address}{New York, NY, USA}, Article
  \bibinfo{articleno}{211}, \bibinfo{numpages}{22}~pages.
\newblock
\showISBNx{9798400703300}
\urldef\tempurl%
\url{https://doi.org/10.1145/3613904.3642730}
\showDOI{\tempurl}


\bibitem[Sharif et~al\mbox{.}(2021)]%
        {sharif_understanding_2021}
\bibfield{author}{\bibinfo{person}{Ather Sharif},
  \bibinfo{person}{Sanjana~Shivani Chintalapati}, \bibinfo{person}{Jacob~O.
  Wobbrock}, {and} \bibinfo{person}{Katharina Reinecke}.}
  \bibinfo{year}{2021}\natexlab{}.
\newblock \showarticletitle{Understanding {Screen}-{Reader} {Users}’
  {Experiences} with {Online} {Data} {Visualizations}}. In
  \bibinfo{booktitle}{\emph{Proceedings of the 23rd {International} {ACM}
  {SIGACCESS} {Conference} on {Computers} and {Accessibility}}}.
  \bibinfo{publisher}{ACM}, \bibinfo{address}{New York, NY, USA},
  \bibinfo{pages}{1--16}.
\newblock
\showISBNx{978-1-4503-8306-6}
\urldef\tempurl%
\url{https://doi.org/10.1145/3441852.3471202}
\showDOI{\tempurl}


\bibitem[Sharif et~al\mbox{.}(2022a)]%
        {sharif_what_2022}
\bibfield{author}{\bibinfo{person}{Ather Sharif}, \bibinfo{person}{Olivia~H.
  Wang}, {and} \bibinfo{person}{Alida~T. Muongchan}.}
  \bibinfo{year}{2022}\natexlab{a}.
\newblock \showarticletitle{“{What} {Makes} {Sonification}
  {User}-{Friendly}?” {Exploring} {Usability} and {User}-{Friendliness} of
  {Sonified} {Responses}}. In \bibinfo{booktitle}{\emph{Proceedings of the 24th
  {International} {ACM} {SIGACCESS} {Conference} on {Computers} and
  {Accessibility}}}. \bibinfo{publisher}{ACM}, \bibinfo{address}{New York, NY,
  USA}, \bibinfo{pages}{1--5}.
\newblock
\showISBNx{978-1-4503-9258-7}
\urldef\tempurl%
\url{https://doi.org/10.1145/3517428.3550360}
\showDOI{\tempurl}


\bibitem[Sharif et~al\mbox{.}(2022b)]%
        {sharif_voxlens_2022}
\bibfield{author}{\bibinfo{person}{Ather Sharif}, \bibinfo{person}{Olivia~H.
  Wang}, \bibinfo{person}{Alida~T. Muongchan}, \bibinfo{person}{Katharina
  Reinecke}, {and} \bibinfo{person}{Jacob~O. Wobbrock}.}
  \bibinfo{year}{2022}\natexlab{b}.
\newblock \showarticletitle{{VoxLens}: {Making} {Online} {Data}
  {Visualizations} {Accessible} with an {Interactive} {JavaScript}
  {Plug}-{In}}. In \bibinfo{booktitle}{\emph{{CHI} {Conference} on {Human}
  {Factors} in {Computing} {Systems}}}. \bibinfo{publisher}{ACM},
  \bibinfo{address}{New York, NY, USA}, \bibinfo{pages}{1--19}.
\newblock
\showISBNx{978-1-4503-9157-3}
\urldef\tempurl%
\url{https://doi.org/10.1145/3491102.3517431}
\showDOI{\tempurl}


\bibitem[Sharif et~al\mbox{.}(2023a)]%
        {sharif_understanding_2023}
\bibfield{author}{\bibinfo{person}{Ather Sharif}, \bibinfo{person}{Andrew~M.
  Zhang}, \bibinfo{person}{Katharina Reinecke}, {and} \bibinfo{person}{Jacob~O.
  Wobbrock}.} \bibinfo{year}{2023}\natexlab{a}.
\newblock \showarticletitle{Understanding and {Improving} {Drilled}-{Down}
  {Information} {Extraction} from {Online} {Data} {Visualizations} for
  {Screen}-{Reader} {Users}}. In \bibinfo{booktitle}{\emph{Proceedings of the
  20th {International} {Web} for {All} {Conference}}}.
  \bibinfo{publisher}{ACM}, \bibinfo{address}{New York, NY, USA},
  \bibinfo{pages}{18--31}.
\newblock
\showISBNx{9798400707483}
\urldef\tempurl%
\url{https://doi.org/10.1145/3587281.3587284}
\showDOI{\tempurl}


\bibitem[Sharif et~al\mbox{.}(2023b)]%
        {sharif_conveying_2023}
\bibfield{author}{\bibinfo{person}{Ather Sharif}, \bibinfo{person}{Ruican
  Zhong}, {and} \bibinfo{person}{Yadi Wang}.} \bibinfo{year}{2023}\natexlab{b}.
\newblock \showarticletitle{Conveying {Uncertainty} in {Data} {Visualizations}
  to {Screen}-{Reader} {Users} {Through} {Non}-{Visual} {Means}}. In
  \bibinfo{booktitle}{\emph{The 25th {International} {ACM} {SIGACCESS}
  {Conference} on {Computers} and {Accessibility}}}. \bibinfo{publisher}{ACM},
  \bibinfo{address}{New York, NY, USA}, \bibinfo{pages}{1--6}.
\newblock
\showISBNx{9798400702204}
\urldef\tempurl%
\url{https://doi.org/10.1145/3597638.3614502}
\showDOI{\tempurl}


\bibitem[Shi et~al\mbox{.}(2019)]%
        {shi_designing_2019}
\bibfield{author}{\bibinfo{person}{Lei Shi}, \bibinfo{person}{Holly Lawson},
  \bibinfo{person}{Zhuohao Zhang}, {and} \bibinfo{person}{Shiri Azenkot}.}
  \bibinfo{year}{2019}\natexlab{}.
\newblock \showarticletitle{Designing {Interactive} {3D} {Printed} {Models}
  with {Teachers} of the {Visually} {Impaired}}. In
  \bibinfo{booktitle}{\emph{Proceedings of the {CHI} {Conference} on {Human}
  {Factors} in {Computing} {Systems}}}. \bibinfo{publisher}{ACM},
  \bibinfo{address}{New York, NY, USA}, \bibinfo{pages}{1--14}.
\newblock
\showISBNx{978-1-4503-5970-2}
\urldef\tempurl%
\url{https://doi.org/10.1145/3290605.3300427}
\showDOI{\tempurl}


\bibitem[Sreevalsan-Nair et~al\mbox{.}(2020)]%
        {sreevalsan-nair_tensor_2020}
\bibfield{author}{\bibinfo{person}{Jaya Sreevalsan-Nair},
  \bibinfo{person}{Komal Dadhich}, {and} \bibinfo{person}{Siri~Chandana
  Daggubati}.} \bibinfo{year}{2020}\natexlab{}.
\newblock \bibinfo{title}{Tensor {Fields} for {Data} {Extraction} from {Chart}
  {Images}: {Bar} {Charts} and {Scatter} {Plots}}.
\newblock
\newblock
\urldef\tempurl%
\url{http://arxiv.org/abs/2010.02319}
\showURL{%
\tempurl}


\bibitem[Thompson et~al\mbox{.}(2023)]%
        {thompson_chart_2023}
\bibfield{author}{\bibinfo{person}{John~R Thompson}, \bibinfo{person}{Jesse~J
  Martinez}, \bibinfo{person}{Alper Sarikaya}, \bibinfo{person}{Edward
  Cutrell}, {and} \bibinfo{person}{Bongshin Lee}.}
  \bibinfo{year}{2023}\natexlab{}.
\newblock \showarticletitle{Chart {Reader}: {Accessible} {Visualization}
  {Experiences} {Designed} with {Screen} {Reader} {Users}}. In
  \bibinfo{booktitle}{\emph{Proceedings of the {CHI} {Conference} on {Human}
  {Factors} in {Computing} {Systems}}} \emph{(\bibinfo{series}{CHI '23})}.
  \bibinfo{publisher}{ACM}, \bibinfo{address}{New York, NY, USA},
  \bibinfo{pages}{1--18}.
\newblock
\showISBNx{978-1-4503-9421-5}
\urldef\tempurl%
\url{https://doi.org/10.1145/3544548.3581186}
\showDOI{\tempurl}


\bibitem[Van~Rooyen et~al\mbox{.}(2024)]%
        {van_rooyen_shape2vibe_2024}
\bibfield{author}{\bibinfo{person}{Xavière Van~Rooyen}, \bibinfo{person}{Gijs
  Huisman}, \bibinfo{person}{Myrthe~A. Plaisier}, {and}
  \bibinfo{person}{Sylvia~C. Pont}.} \bibinfo{year}{2024}\natexlab{}.
\newblock \showarticletitle{{Shape2Vibe}: {A} {Tangible} {Tool} for
  {Vibrotactile} {Co}-{Design} with {People} with {Deafblindness}}. In
  \bibinfo{booktitle}{\emph{Proceedings of the {Eighteenth} {International}
  {Conference} on {Tangible}, {Embedded}, and {Embodied} {Interaction}}}
  \emph{(\bibinfo{series}{{TEI} '24})}. \bibinfo{publisher}{ACM},
  \bibinfo{address}{New York, NY, USA}, \bibinfo{pages}{1--6}.
\newblock
\showISBNx{9798400704024}
\urldef\tempurl%
\url{https://doi.org/10.1145/3623509.3635264}
\showDOI{\tempurl}


\bibitem[W3C(2024)]%
        {wai-aria}
\bibfield{author}{\bibinfo{person}{W3C}.} \bibinfo{year}{2024}\natexlab{}.
\newblock \bibinfo{title}{Accessible {Rich} {Internet} {Applications}
  ({WAI}-{ARIA}) 1.2}.
\newblock
\newblock
\urldef\tempurl%
\url{https://www.w3.org/TR/wai-aria/}
\showURL{%
\tempurl}


\bibitem[White et~al\mbox{.}(2008)]%
        {white_toward_2008}
\bibfield{author}{\bibinfo{person}{Gareth~R. White}, \bibinfo{person}{Geraldine
  Fitzpatrick}, {and} \bibinfo{person}{Graham McAllister}.}
  \bibinfo{year}{2008}\natexlab{}.
\newblock \showarticletitle{Toward accessible {3D} virtual environments for the
  blind and visually impaired}. In \bibinfo{booktitle}{\emph{Proceedings of the
  3rd international conference on {Digital} {Interactive} {Media} in
  {Entertainment} and {Arts}}} \emph{(\bibinfo{series}{{DIMEA} '08})}.
  \bibinfo{publisher}{ACM}, \bibinfo{address}{New York, NY, USA},
  \bibinfo{pages}{134--141}.
\newblock
\showISBNx{978-1-60558-248-1}
\urldef\tempurl%
\url{https://doi.org/10.1145/1413634.1413663}
\showDOI{\tempurl}


\bibitem[Wobbrock(2003)]%
        {wobbrock2003benefits}
\bibfield{author}{\bibinfo{person}{Jacob Wobbrock}.}
  \bibinfo{year}{2003}\natexlab{}.
\newblock \showarticletitle{The benefits of physical edges in gesture-making:
  Empirical support for an edge-based unistroke alphabet}. In
  \bibinfo{booktitle}{\emph{CHI'03 Extended Abstracts on Human Factors in
  Computing Systems}}. \bibinfo{pages}{942--943}.
\newblock


\bibitem[Zhang et~al\mbox{.}(2023)]%
        {zhang_developing_2023}
\bibfield{author}{\bibinfo{person}{Zhuohao Zhang}, \bibinfo{person}{Gene S-H
  Kim}, {and} \bibinfo{person}{Jacob~O. Wobbrock}.}
  \bibinfo{year}{2023}\natexlab{}.
\newblock \showarticletitle{Developing and {Deploying} a {Real}-{World}
  {Solution} for {Accessible} {Slide} {Reading} and {Authoring} for {Blind}
  {Users}}. In \bibinfo{booktitle}{\emph{Proceedings of the 25th
  {International} {ACM} {SIGACCESS} {Conference} on {Computers} and
  {Accessibility}}} \emph{(\bibinfo{series}{{ASSETS} '23})}.
  \bibinfo{publisher}{ACM}, \bibinfo{address}{New York, NY, USA},
  \bibinfo{pages}{1--15}.
\newblock
\showISBNx{9798400702204}
\urldef\tempurl%
\url{https://doi.org/10.1145/3597638.3608418}
\showDOI{\tempurl}


\bibitem[Zhang and Wobbrock(2023)]%
        {zhang_a11yboard_2023}
\bibfield{author}{\bibinfo{person}{Zhuohao~(Jerry) Zhang} {and}
  \bibinfo{person}{Jacob~O. Wobbrock}.} \bibinfo{year}{2023}\natexlab{}.
\newblock \showarticletitle{{A11yBoard}: {Making} {Digital} {Artboards}
  {Accessible} to {Blind} and {Low}-{Vision} {Users}}. In
  \bibinfo{booktitle}{\emph{Proceedings of the {CHI} {Conference} on {Human}
  {Factors} in {Computing} {Systems}}}. \bibinfo{publisher}{ACM},
  \bibinfo{address}{New York, NY, USA}, \bibinfo{pages}{1--17}.
\newblock
\showISBNx{978-1-4503-9421-5}
\urldef\tempurl%
\url{https://doi.org/10.1145/3544548.3580655}
\showDOI{\tempurl}


\bibitem[Zhao et~al\mbox{.}(2008)]%
        {zhao_data_2008}
\bibfield{author}{\bibinfo{person}{Haixia Zhao}, \bibinfo{person}{Catherine
  Plaisant}, \bibinfo{person}{Ben Shneiderman}, {and} \bibinfo{person}{Jonathan
  Lazar}.} \bibinfo{year}{2008}\natexlab{}.
\newblock \showarticletitle{Data {Sonification} for {Users} with {Visual}
  {Impairment}: {A} {Case} {Study} with {Georeferenced} {Data}}.
\newblock \bibinfo{journal}{\emph{ACM Transactions on Computer-Human
  Interaction}} \bibinfo{volume}{15}, \bibinfo{number}{1} (\bibinfo{date}{May}
  \bibinfo{year}{2008}), \bibinfo{pages}{4:1--4:28}.
\newblock
\showISSN{1073-0516}
\urldef\tempurl%
\url{https://doi.org/10.1145/1352782.1352786}
\showDOI{\tempurl}


\bibitem[Zhao et~al\mbox{.}(2024)]%
        {zhao_tada_2024}
\bibfield{author}{\bibinfo{person}{Yichun Zhao}, \bibinfo{person}{Miguel~A
  Nacenta}, \bibinfo{person}{Mahadeo~A. Sukhai}, {and} \bibinfo{person}{Sowmya
  Somanath}.} \bibinfo{year}{2024}\natexlab{}.
\newblock \showarticletitle{TADA: Making Node-link Diagrams Accessible to Blind
  and Low-Vision People}. In \bibinfo{booktitle}{\emph{Proceedings of the CHI
  Conference on Human Factors in Computing Systems}}
  \emph{(\bibinfo{series}{CHI '24})}. \bibinfo{publisher}{ACM},
  \bibinfo{address}{New York, NY, USA}, Article \bibinfo{articleno}{45},
  \bibinfo{numpages}{20}~pages.
\newblock
\showISBNx{9798400703300}
\urldef\tempurl%
\url{https://doi.org/10.1145/3613904.3642222}
\showDOI{\tempurl}


\bibitem[Zong et~al\mbox{.}(2022)]%
        {zong_rich_2022}
\bibfield{author}{\bibinfo{person}{Jonathan Zong}, \bibinfo{person}{Crystal
  Lee}, \bibinfo{person}{Alan Lundgard}, \bibinfo{person}{JiWoong Jang},
  \bibinfo{person}{Daniel Hajas}, {and} \bibinfo{person}{Arvind Satyanarayan}.}
  \bibinfo{year}{2022}\natexlab{}.
\newblock \showarticletitle{Rich {Screen} {Reader} {Experiences} for
  {Accessible} {Data} {Visualization}}.
\newblock \bibinfo{journal}{\emph{Computer Graphics Forum}}
  \bibinfo{volume}{41}, \bibinfo{number}{3} (\bibinfo{date}{June}
  \bibinfo{year}{2022}), \bibinfo{pages}{15--27}.
\newblock
\showISSN{0167-7055, 1467-8659}
\urldef\tempurl%
\url{https://doi.org/10.1111/cgf.14519}
\showDOI{\tempurl}


\bibitem[Zong et~al\mbox{.}(2024)]%
        {zong_umwelt_2024}
\bibfield{author}{\bibinfo{person}{Jonathan Zong}, \bibinfo{person}{Isabella
  Pedraza~Pineros}, \bibinfo{person}{Mengzhu~(Katie) Chen},
  \bibinfo{person}{Daniel Hajas}, {and} \bibinfo{person}{Arvind Satyanarayan}.}
  \bibinfo{year}{2024}\natexlab{}.
\newblock \showarticletitle{Umwelt: Accessible Structured Editing of
  Multi-Modal Data Representations}. In \bibinfo{booktitle}{\emph{Proceedings
  of the CHI Conference on Human Factors in Computing Systems}}
  \emph{(\bibinfo{series}{CHI '24})}. \bibinfo{publisher}{ACM},
  \bibinfo{address}{New York, NY, USA}, Article \bibinfo{articleno}{46},
  \bibinfo{numpages}{20}~pages.
\newblock
\showISBNx{9798400703300}
\urldef\tempurl%
\url{https://doi.org/10.1145/3613904.3641996}
\showDOI{\tempurl}


\end{thebibliography}

\setcounter{section}{0}

\section*{Appendix}

\begin{algorithm}
\caption{The implementation of the dynamic scanning radius feature}
\label{alg}
\begin{algorithmic}[1]
    \State $indicesWithinRadius \gets []$
    \State $radiusCoverDistance \gets$ number of data points the dynamic radius should cover
    \State $minRad, maxRad \gets$ the min and max radius allowed to avoid edge cases

    \Function{touchMoved}{$position$}
    \State $distances, hitIndices \gets []$
    \ForAll{$dp$ in $datapoints$}
        \State $distances.\text{insert}(\text{euclideanDistance}(dp, position))$
    \EndFor
    \State $sortedDistances, indices \gets \text{sort}(distances)$
    \State $rawRadius \gets sortedDistances[radiusCoverDistance-1]$ 
    \State $adjustedRadius \gets \text{clamp}(rawRadius, minRad, maxRad)$
    \For{$i \gets 0$ \textbf{to} $length(sortedDistances) - 1$}
        \If{$sortedDistances[i] \leq adjustedRadius$}
            \State $hitIndices.\text{insert}(indices[i])$
        \EndIf
    \EndFor
    \State $newlyHitTargets \gets$ indices not in $indicesWithinRadius$ but in $hitIndices$
    \State $indicesWithinRadius \gets hitIndices$
    \If{$\text{length}(newlyHitTargets) > 0$}
        \State Fire haptic vibration for $\text{length}(newlyHitTargets)$ times
    \EndIf
    \EndFunction

\end{algorithmic}
\end{algorithm}

\end{document}